\documentclass[journal]{IEEEtran}
\usepackage{amsmath,amsfonts}
\usepackage{array}
\usepackage{textcomp}
\usepackage{stfloats}
\usepackage{url}
\usepackage{verbatim}
\usepackage{graphicx}
\usepackage{balance}
\usepackage{cite}
\usepackage{cleveref}
\usepackage{dsfont}
\usepackage{color}
\usepackage[caption=false,font=footnotesize]{subfig}
\usepackage{mathtools, cuted}
\usepackage{makecell}

\usepackage{microtype}
\usepackage{multicol}
\usepackage{booktabs}

\usepackage{amsmath}
\usepackage{amssymb}
\usepackage{mathtools}
\usepackage{amsthm}

\newcommand{\imp}{{\textrm{imp}}}
\newcommand{\offset}{{\textrm{offset}}}

\newcommand{\CPP}{{\textrm{CPP}}}
\newcommand{\PSNR}{{\textrm{PSNR}}}
\newcommand{\SNR}{{\textrm{SNR}}}
\newcommand{\MSE}{{\textrm{MSE}}}
\newcommand{\MAXX}{{\textrm{MAX}}}
\newcommand{\SSIM}{{\textrm{SSIM}}}
\newcommand{\FA}{{\textrm{FA}}}
\newcommand{\distortion}{{\textrm{distortion}}}
\newcommand{\inn}{{\textrm{in}}}
\newcommand{\out}{{\textrm{out}}}
\newcommand{\tr}{{\textrm{train}}}
\newcommand{\soft}{{\textrm{soft}}}
\newcommand{\hard}{{\textrm{hard}}}
\newcommand{\ST}{{\textrm{ST}}}
\newcommand{\onehot}{{\textrm{one\_hot}}}
\newcommand{\stopg}{{\textrm{stop\_grad}}}
\newcommand{\selfa}{{\textrm{self\_attention}}}
\newcommand{\defo}{{\textrm{deformation}}}
\newcommand{\mean}{{\textrm{mean}}}
\newcommand{\std}{{\textrm{std}}}
\DeclareMathOperator*{\argmax}{arg\,max}
\newcommand{\norm}[1]{\left\lVert#1\right\rVert}
\begin{document}

\title{Feature Importance-Aware Deep Joint Source-Channel Coding for Computationally Efficient and Adjustable Image Transmission}
\author{
  Hansung Choi and Daewon Seo\IEEEmembership{}
    \thanks{The authors are with the Department of Electrical Engineering and Computer Science, Daegu Gyeongbuk Institute of Science and Technology (DGIST), Daegu 42988, South Korea (e-mail: \{hansungchoi, dwseo\}@dgist.ac.kr).}
}

\maketitle

\begin{abstract}
Recent advances in deep learning-based joint source–channel coding (deepJSCC) have substantially improved communication performance, but their growing computational cost hinders practical deployment. Moreover, certain applications require the ability to dynamically adapt computational complexity. To address these issues, we propose a Feature Importance-Aware deepJSCC (FAJSCC) model for image transmission that is both computationally efficient and adjustable. FAJSCC employs axis-dimension specialized computation, which performs efficient operations individually for each spatial and channel axis, significantly reducing computational cost while representing features effectively. It further incorporates selective deformable self-attention, which applies self-attention only to selected and adaptively adjusted features, leveraging the importance and relations of input features to efficiently capture complex feature correlations. Another key feature of FAJSCC is that the number of selected important areas can be controlled separately by the encoder and the decoder, depending on the available computational budget. It makes FAJSCC the first deepJSCC architecture to allow independent adjustment of encoder and decoder complexity within a single trained model. Experimental results show that FAJSCC achieves superior image transmission performance under various channel conditions while requiring less computational complexity than recent state-of-the-art models. Furthermore, experiments independently varying the encoder and decoder's computational resources reveal, for the first time in the deepJSCC literature, that understanding the meaning of noisy features in the decoder demands the greatest computational cost. The code is publicly available at github.com/hansung-choi/FAJSCCv2.
\end{abstract}


\section{Introduction} \label{sec:introduction}

Advances in high-performance embedded hardware and next-generation wireless communication technologies such as 6G have enabled the widespread deployment of IoT devices across diverse application domains, including smart healthcare, smart manufacturing, and smart cities. In particular, vision IoT devices, such as wireless surveillance cameras and UAVs, provide important visual information for security and safety monitoring, including real-time intrusion detection and inspection of hazardous areas. However, the rapid growth of IoT devices has increased pressure on available bandwidth. Especially, vision IoT devices require substantial bandwidth to transmit visual data, which are inherently more complex and contain a much larger amount of information, thereby requiring significantly more bits than other modalities such as text and speech. Therefore, the recent communication systems need high-quality vision data transmission methods under limited bandwidth constraints.

Although conventional separation-based digital communication systems have undergone rapid improvements in both source coding (e.g., JPEG, JPEG2000, PNG, BPG, and AVIF) and channel coding (e.g., Turbo, LDPC, and polar codes) over several decades, further manual improvements to these rule-based methods have become increasingly difficult in satisfying recent communication requirements. In the other side, it is known that the joint source-channel coding (JSCC) design strictly outperforms the separation-based design in several scenarios, such as finite block length~\cite{ho2013separation} and sending correlated data over multi-user settings~\cite{cover1980multiple}. Early analytic research on the JSCC was limited to simple data distributions, like Gaussian cases, due to the tractability, and hence could not deal with real data, such as images~\cite{hekland2009shannon,hu2011analog}.

One of the breakthroughs in advancing the JSCC beyond its early results is the use of deep learning, known as deepJSCC, which has shown significant performance improvement for image transmission~\cite{bourtsoulatze2019deep} and become close to satisfying recent communication requirements. Since the introduction of deepJSCC, extensive efforts have been made, leading to great success in terms of communication performance. For instance, deepJSCC models have shown better data reconstruction performance compared to separation-based systems for simple point-to-point additive white Gaussian noise (AWGN) channels~\cite{bourtsoulatze2019deep,zhang2023predictive,yang2024swinjscc,cheng2025tcc,zhang2025snr}, various fading channels ~\cite{liu2025semantic,xu2021wireless,wu2024mambajscc1,yilmaz2024high,zhang2025semantics,wu2024cddm}, and several bandwidth settings~\cite{kurka2021bandwidth,bian2023deepjscc,yang2022deep} as well as diverse network configurations such as MIMO~\cite{wu2024deep} and relay networks~\cite{bian2025process}. Recent vision-related semantic communication can also be viewed as a variant of deepJSCC designed for downstream tasks, such as image retrieval~\cite{lo2023collaborative}, segmentation~\cite{lv2024importance}, and multiple tasks~\cite{yu2025multi,liu2024rate}.

Despite its success in communication performance, existing deepJSCC research is increasingly suffering from growing computational complexity incurred by large neural networks, along with associated problems like high power consumption, hardware costs, and latency, which prohibit its practical deployment, especially for IoT devices. To reduce such computational burden, one common approach is to apply model compression techniques. For instance, pruning~\cite{jankowski2020joint,zhang2024engineering,zhang2024lightweight} and low-rank decomposition~\cite{xu2022low} compress trained models, successfully reducing the number of parameters (i.e., computational load) at the expense of performance degradation and extra training efforts.

Beyond such compression-based strategies, another line of work focuses on designing lightweight models~\cite{yu2025novel,niu2025multimodal,huang2025progressive,chen2024lightweight,jia2023lightweight}. For example, reducing the intermediate deepJSCC feature channel size before passing through computationally intensive blocks~\cite{jia2023lightweight} or performing computations only on a fixed subset of feature channels~\cite{yu2025novel,huang2025progressive} can lower the computational load. In addition, approaches that remove the least important features have been proposed~\cite{niu2025multimodal,zhang2024unified}. While these methods alleviate computational burden without requiring extra training, reducing or deleting features and restricting computations to limited channel regions degrade image transmission performance. Therefore, there is an urgent need for deepJSCC designs that can improve transmission performance while reducing computational cost compared to the current state-of-the-art models.

Another important function that is often required in applications is the ability to dynamically control computational complexity within a single model. For instance, in vision monitoring applications of IoT devices, it is essential to transmit images of large areas at low computational complexity (i.e., low resolution) to ensure minimal delay and power consumption for long-term monitoring, while transmitting at high computational complexity (i.e., high resolution) when detailed analysis is needed. To address the varying computational needs at the application levels, it is essential to dynamically manage the tradeoff between computational complexity and data communication accuracy.

Although dynamic inference mechanisms such as routing, token pruning, and early-exit have been widely studied to adjust computational complexity in deep learning~\cite{shazeer2017outrageously,rao2021dynamicvit,teerapittayanon2016branchynet}, they are not well suited to communication scenarios. Routing-based approaches require multiple models to be stored, leading to large memory overhead, which is impractical for memory-constrained IoT devices~\cite{lin2024computation}. Token pruning-based designs reduce computational cost by discarding less important features~\cite{rao2021dynamicvit}, but this strategy is unsuitable for image transmission where all visual information must be preserved, and it also limits computational adaptability at the receiver side~\cite{niu2025multimodal,zhang2024unified}. Early-exit based designs such as DD-JSCC~\cite{raha2025dd} adjust complexity by skipping intermediate layers, but the degree of layer skipping must be synchronized between the transmitter and receiver due to the dependency of feature distributions across layers. Consequently, the computational complexity is constrained by the side with stricter resource limitations and cannot be independently adjusted when the transmitter and receiver have heterogeneous computational capabilities. These limitations indicate that directly adopting existing dynamic inference frameworks is insufficient for deepJSCC systems, motivating the need for a communication-oriented design that enables independent computational resource adaptation within a single model.

 To simultaneously achieve computational efficiency and enable independent dynamic control of computational complexity at both the transmitter and receiver, we propose a feature importance-aware deepJSCC (FAJSCC) framework. FAJSCC employs dimension-specialized computation strategies that enhance computational efficiency while achieving high transmission performance. The first strategy is axis-dimension specialized computation. A standard convolution over all channels, height, and width can be factorized into a depthwise convolution over the spatial dimensions (height and width) and a pointwise convolution over the channel dimension. This decomposition substantially reduces the number of operations: Instead of jointly processing all input and output channels with multi-channel kernels, the depthwise convolution handles each channel separately with each single-channel kernel, and the pointwise convolution mixes channels with 1×1 kernels. By exploiting this separation, FAJSCC achieves significantly lower computational cost while maintaining the feature processing capacity of standard convolution. To further enhance these operations, FAJSCC integrates spatial and channel attention mechanisms. The spatial attention emphasizes key features along the spatial axes, whereas the channel attention highlights important channels~\cite{woo2018cbam}. Accordingly, spatial attention is applied before depth-wise convolution and channel attention before point-wise convolution, maximizing feature processing for each axis. These lightweight attention operations exhibit linear computational complexity with respect to the input feature sizes. Our experiments demonstrate that even using only axis-dimension specialized computation can outperform the current state-of-the-art SwinJSCC~\cite{yang2024swinjscc} while using only half of the computational resources.

The second strategy is important-dimension specialized computation, which captures relations of complex features that axis-dimension operations may miss. To this end, FAJSCC employs a modified window self-attention mechanism~\cite{liu2021swin}. The window self-attention mechanism shows highly effective feature processing ability by capturing relations of complex features for a given fixed window area in various semantic communication scenarios~\cite{zhang2025semantic,wu2024cddm,yang2024swinjscc,cheng2025tcc,wang2025task}. However, conventional window self-attention has two limitations: it cannot capture highly related features across different windows, and analyzing features of every window is computationally expensive. FAJSCC addresses these issues through a new attention mechanism, which we call selective deformable self-attention. The deformable self-attention adjusts self-attention regions based on the window and offsets computed from the input feature relation, enabling adaptive feature analysis beyond the fixed window. 

Importantly, FAJSCC applies this deformable self-attention selectively based on feature importance. Some features are more informative (e.g., main objects), while others are less informative (e.g., background regions)~\cite{cheng2020learned}. Less informative features can be processed sufficiently by the preceding axis-dimension specialized operations alone. Moreover, simple images require only minimal computation to achieve high task performance~\cite{kong2021classsr,hang2025rate}. Inspired by these findings, FAJSCC applies this deformable self-attention only to the most important features rather than to all features. This selective enhancement strategy greatly reduces computational cost compared with applying self-attention across all windows. A portion of the saved computation is then reinvested to increase intermediate feature channel sizes, leading to richer representations. Although this reinvestment slightly raises the complexity, the increase is much smaller than the computational savings, so the overall cost remains significantly lower than applying self-attention to the entire feature set. As a result, FAJSCC achieves higher transmission performance by enhancing important feature processing while still maintaining reduced overall complexity. Unlike previous lightweight deepJSCC methods that reduce or remove features~\cite{jia2023lightweight,niu2025multimodal,zhang2024unified}, which degrade performance, FAJSCC enlarges and selectively refines important features, which increase performance while reducing computational burden. Moreover, due to our proposed selective feature enhancement framework, the transmitter and the receiver of FAJSCC can adjust their computational complexity independently, unlike previous resource-adjustable communication methods where resource adjustments are constrained by the other side~\cite{niu2025multimodal,zhang2024unified,raha2025dd}.

Finally, to further maximize efficiency, FAJSCC introduces a novel attention family tree that simultaneously extracts spatial and channel attention features, offsets, and feature importance while eliminating overlapping computations. Deformation offsets are derived from spatial-wise feature information, and spatial attention is generated from spatial-wise feature importance, which itself can be computed from the same spatial-wise feature information. Thus, redundant pre-processing steps can be removed by reusing spatial-wise feature information to get spatial-wise feature importance. Moreover, spatial-wise feature importance serves a dual purpose: identifying feature importance for selective self-attention and producing spatial attention features. Consequently, attentions, offsets, and feature importance are obtained efficiently without incurring significant computational overhead. Ablation studies confirm that the attention family tree achieves comparable or superior performance while reducing computational cost relative to extracting these features separately.

To adapt to varying computational budgets, we introduce the importance ratio $\gamma$, representing the fraction of features selected as important. When $\gamma = 1$, all features are processed via deformable self-attention, whereas $\gamma = 0$ means none are processed. For $0<\gamma<1$, the top $\gamma$ fraction of features, ranked by their measured importance, are selected and processed through deformable self-attention. By adjusting the importance ratio, FAJSCC enables controllable computational complexity within a single model. Previous resource-adjustable deepJSCC methods achieve complexity control by deleting features~\cite{niu2025multimodal,zhang2024unified}, which can lead to significant performance degradation due to information loss. In contrast, FAJSCC maintains high performance because its selective feature enhancement strategy preserves all information.

Moreover, FAJSCC decouples computational complexity from the model architecture, allowing first independent evaluation of the impact of encoder and decoder complexity on performance. It is important to note that typical previous deepJSCC models have fixed and symmetric computational complexity at encoder and decoder~\cite{bourtsoulatze2019deep,yang2024swinjscc,wu2024mambajscc1} or the transceiver's computational complexity adjustments are constrained by the counterpart communication side~\cite{niu2025multimodal,zhang2024unified,raha2025dd}. As a result, such a discussion has not been possible in previous models. Recall that the encoder compresses images and adds redundancy to protect data against channel noise, while the decoder interprets the received noisy signal and (re)generates the transmitted images. A natural question arises: Which component requires the most computational complexity, and under what conditions? By independently varying the importance ratios for the encoder and decoder across different signal-to-noise ratios (SNRs), we draw two key conclusions: 1) the decoder requires higher computational complexity than the encoder, and 2) the decoder's noisy-signal perception demands even greater complexity when the SNR is low. To the best of our knowledge, this is the first detailed analysis of computational complexity in deepJSCC.

As a consequence of our design, our FAJSCC achieves higher image transmission performance than other deepJSCC models, even with lower computational complexity compared to the previous state-of-the-art (SOTA) model~\cite{yang2024swinjscc}. Our FAJSCC also shows higher efficiency than the recent light-weight model~\cite{yu2025novel}. Our main contributions can be summarized as follows.
\begin{itemize}
    \item \textbf{Efficient Computation:} We propose axis-dimension specialized computation and selective deformable self-attention to significantly reduce computational cost while maintaining high transmission performance. By processing only the most important features with deformable self-attention and efficiently reusing attention-related information through the attention family tree, FAJSCC achieves superior image reconstruction compared to existing SOTA models with lower computational requirements.
    
    \item \textbf{Adjustable Computation:} We introduce the importance ratio to control the fraction of features processed by deformable self-attention, enabling FAJSCC to separately adjust the computational complexities of the encoder and the decoder. Unlike FAJSCC, previous resource-adjustable deep learning based communications are constrained by the counterpart communication side~\cite{niu2025multimodal,zhang2024unified,raha2025dd}.
    
    \item \textbf{Computational Complexity vs.~Performance:} By varying the computational resources allocated to the encoder and decoder across different SNRs, we can identify which function is most sensitive to the variation of computational complexity and has the greatest impact on performance. Our experiment suggests that the decoder's noisy-signal perception is the most critical factor influencing performance, providing insight for future advancements aimed at maximizing the efficiency of deepJSCC. Note that FAJSCC is the first deepJSCC model that enables this analysis without modifying the architecture or scaling the model.
\end{itemize}

The remaining parts are organized as follows. In Section~\ref{sec:system model}, we formally introduce the communication system considered in this paper. In Section~\ref{sec:FAJSCC framework}, we present our FAJSCC structure in details. In Section~\ref{sec:experiment}, the performance of FAJSCC is demonstrated with several deepJSCC baselines under various communication conditions. Finally, we conclude our paper in Section~\ref{sec:conclusion}.

\begin{figure}[t]
    \centering
    \includegraphics[width=0.9\linewidth]{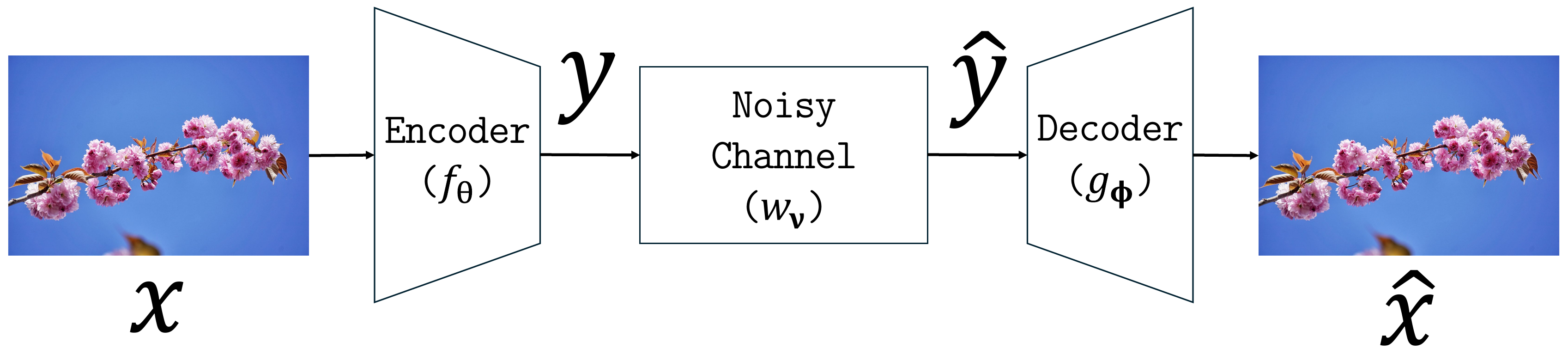}
    \caption{A point-to-point deepJSCC communication system.}\label{fig:deepJSCCsystem}
    \vspace{-0.15in}
\end{figure}

\section{Problem Formulation}
\label{sec:system model}

We consider a point-to-point image transmission system using deepJSCC over a noisy channel, as shown in Figure~\ref{fig:deepJSCCsystem}. The deepJSCC encoder $f_{\theta}$, with model parameter $\theta$, maps a source image $\mathbf{x} \in \mathbb{R}^{H \times W \times 3}$ to a channel input $\mathbf{y} \in \mathbb{C}^{k}$, where $H, W$, and $3$ represent the height, width, and RGB color channels of the source image $\mathbf{x}$, respectively. Here, $k$ is the number of transmitted symbols, also called the channel bandwidth. The ratio of the bandwidth to the total number of RGB pixels is particularly called the bandwidth ratio or the channel per pixel (CPP), i.e.,
\begin{align}
    \CPP := \frac{k}{3HW}.
\end{align}
We assume that the average power of $\mathbf{y}$ does not exceed the power constraint $P$, i.e., $\frac{1}{k} \norm{\mathbf{y}}_{2}^{2} \le P$. For simplicity, we set $P=1$ in this paper without loss of generality.

After encoding, the transmitted signal $\mathbf{y}$ is corrupted by the channel noise, and then the decoder receives $\hat{\mathbf{y}} \in \mathbb{C}^k$. The relationship between $y$ and $\hat{y}$ depends on the type of channel. In this paper, we consider two channel models: the additive white Gaussian noise (AWGN) channel and the fast Rayleigh fading channel. In the AWGN channel, we assume that the noise power is $\sigma^2$, i.e., the relationship between the $i$-th symbols of $\mathbf{y}$ and $\hat{\mathbf{y}}$ is given by 
\begin{align}
    \hat{y}_i = y_i + \epsilon_i, ~~ \epsilon_i \sim \mathcal{CN}(0,\sigma^2), ~ i \in [1:k] 
\end{align}
where $\mathcal{CN}(0,\sigma^2)$ is the circularly symmetric complex Gaussian distribution with mean $0$ and variance $\sigma^2$.

In the case of the fast Rayleigh fading channel with noise power $\sigma^2$, the relationship between the $i$-th symbols of $\mathbf{y}$ and $\hat{\mathbf{y}}$ is given by
\begin{align}
    \hat{y}_i = h_i y_i + \epsilon_i, ~~ h_i \sim \mathcal{CN}(0,1), ~ i \in [1:k] 
\end{align}
where $h_i$ represents the Rayleigh fading coefficient. The quality of these channel models is measured by the signal-to-noise ratio (SNR), which is defined in decibels (dB) as follows.
\begin{align}
    \SNR := 10 \log_{10} \left( \frac{P}{\sigma^2} \right) = 10 \log_{10} \left( \frac{1}{\sigma^2} \right) ~ \text{dB} 
\end{align}
since $P=1$ in this paper.

\begin{figure*}[t]
    \centering
    \includegraphics[width=0.7\linewidth]{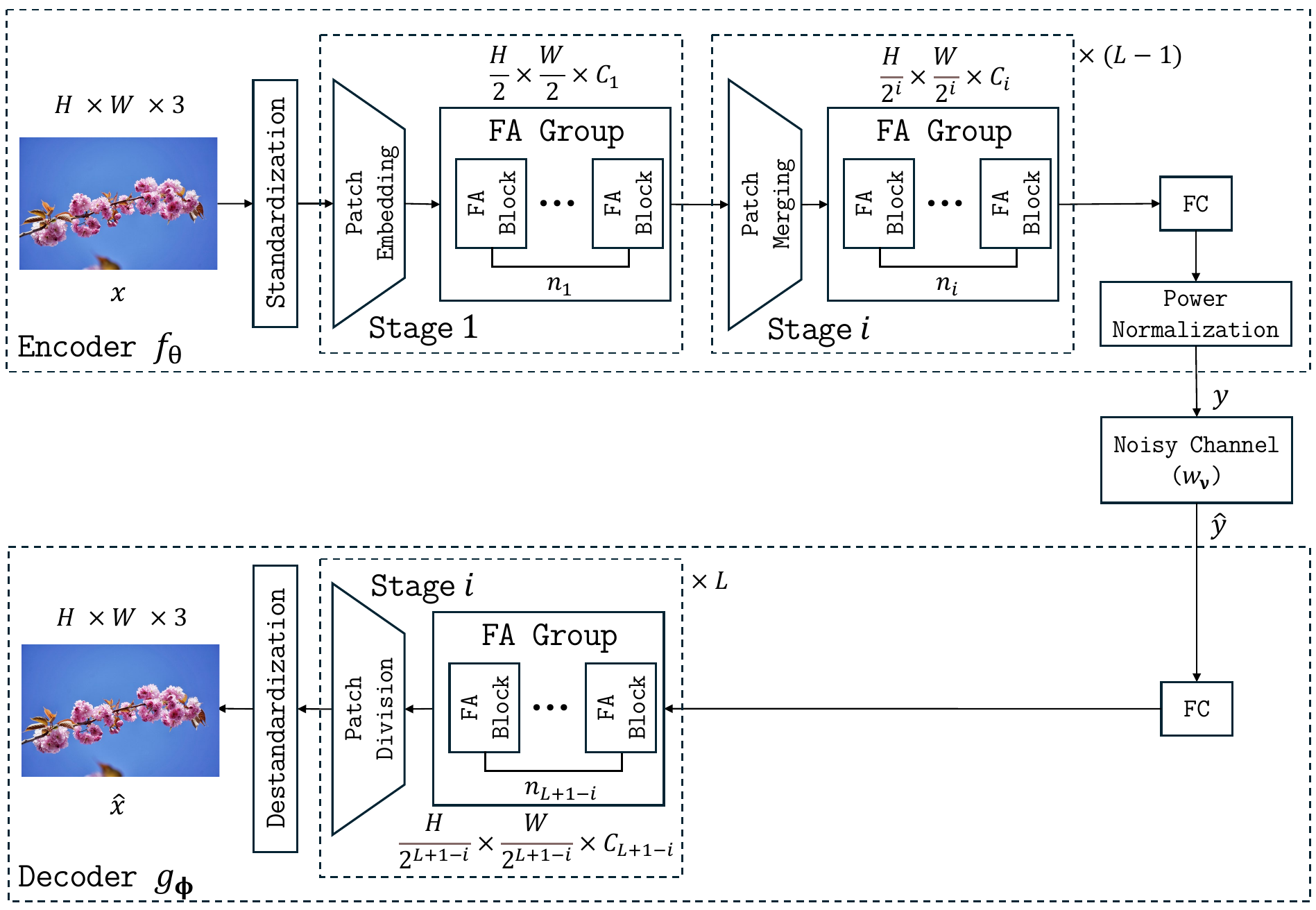}
    \caption{FAJSCC Architecture.}\label{fig:deepJSCCarchitecture}
    \vspace{-0.15in}
\end{figure*}

After receiving $\hat{\mathbf{y}} \in \mathbb{C}^{k}$, for the AWGN channel, the deepJSCC decoder $g_{\phi}$ with parameter $\phi$ maps the received signal $\hat{\mathbf{y}}$ to a reconstructed source image $\hat{\mathbf{x}} \in \mathbb{R}^{H \times W \times 3}$. In case of a fast Rayleigh fading channel, we assume perfect CSI only at the receiver side, i.e., Rayleigh fading coefficients are known to the receiver. We use the equalized signal $\bar{y}$ as decoder input with minimum mean square error equalization (MMSE), i.e., $\bar{\mathbf{y}}=\frac{\mathbf{h}^*}{\mathbf{h}^*\mathbf{h}+\sigma^2}\hat{\mathbf{y}}$. Here $\mathbf{h}^*$ is the complex conjugate of $\mathbf{h}$.

Our goal is to develop a computationally efficient and adjustable deepJSCC model that minimizes the distortion between the original and reconstructed images, $\mathbf{x}$ and $\hat{\mathbf{x}}$, respectively. This objective is particularly important because accurate image reconstruction is central to our target scenarios, such as wireless surveillance and monitoring applications. In this paper, the quality of reconstructed images is measured in two ways. Firstly, peak signal-to-noise ratio (PSNR) measures the amount of pixel-wise distortion, defined in decibel (dB) as follows.
\begin{align}
    \PSNR := 10 \log_{10} \frac{\MAXX^2}{\MSE} ~ \text{dB},
\end{align}
where $\MAXX$ is a pixel's maximum value, which is $255$ in our case (i.e., $8$-bit representation), and MSE is the mean-squared error (MSE) defined by
\begin{align}
    \MSE := \frac{1}{H \times W \times 3} \| \mathbf{x} - \hat{\mathbf{x}} \|_2^2.
\end{align}
Secondly, the structural similarity index measure (SSIM) evaluates image quality based on brightness, contrast, and structure~\cite{wang2004image}. Specifically, SSIM is defined by
\begin{align}
    \SSIM(\mathbf{x},\hat{\mathbf{x}}) := l(\mathbf{x},\hat{\mathbf{x}})^{p} c(\mathbf{x},\hat{\mathbf{x}})^{q} s(\mathbf{x},\hat{\mathbf{x}})^{r},
\end{align}
where hyperparameters $p, q, r$ weight the contributions of brightness similarity comparison $l(\mathbf{x},\hat{\mathbf{x}})$, contrast similarity comparison $c(\mathbf{x},\hat{\mathbf{x}})$, and structural similarity comparison $s(\mathbf{x},\hat{\mathbf{x}})$, respectively. Since SSIM considers multiple aspects that contribute to structural similarities, a higher SSIM value indicates better structure accuracy.

To measure the efficiency of the deepJSCC model, we evaluate computational resource usage in terms of floating-point operations (FLOPs) and memory usage in bytes. FLOPs refer to the number of arithmetic operations involving floating-point numbers, such as additions and multiplications. We also measure latency (ms/image), peak memory (MB), and model storage size (MB). Here, latency is the end-to-end execution time from encoder to decoder, and peak memory is the maximum GPU memory allocation (MB) observed during a single forward pass of the model.

\section{FAJSCC Framework}
\label{sec:FAJSCC framework}
Our overall FAJSCC structure is illustrated in Figure~\ref{fig:deepJSCCarchitecture}. To facilitate deepJSCC training, standardization of the encoder maps pixel values of an image from $[0,255]$ to $[-1,1]$ while destandardization of the decoder reverses this process. Power normalization of the encoder rescales features to fit the power constraint $P=1$. Following previous works~\cite{zhang2025semantic,wu2024cddm,yang2024swinjscc,zhang2025snr}, our FAJSCC uses patch embedding to refine the image at the feature level, patch merging for down-sampling, and patch division for up-sampling. Based on the feature sizes, both the encoder and decoder consist of multiple stages, indexed by $i \in [1:L]$, where $L$ is the total number of stages. The patch merging layer of the $i$-th stage transforms the feature dimension from $\frac{H}{2^{i-1}} \times \frac{W}{2^{i-1}} \times C_{i-1}$ to $\frac{H}{2^{i}} \times \frac{W}{2^{i}} \times C_{i}$. Conversely, the patch division layer in the $i$-th stage transforms the feature dimensions from $\frac{H}{2^{L+1-i}} \times \frac{W}{2^{L+1-i}} \times C_{L+1-i}$ to $\frac{H}{2^{L-i}} \times \frac{W}{2^{L-i}} \times C_{L-i}$. Here, $C_{0}$ is the number of RGB color channels, i.e., $C_{0}=3$. The encoder's fully connected (FC) layer scales input feature channel dimensions to fit the available bandwidth, while the decoder's FC layer adjusts input feature channel dimensions to align the feature channel sizes of the decoder's first stage.

\begin{figure*}[t]
    \centering
    \includegraphics[width=1.0\linewidth]{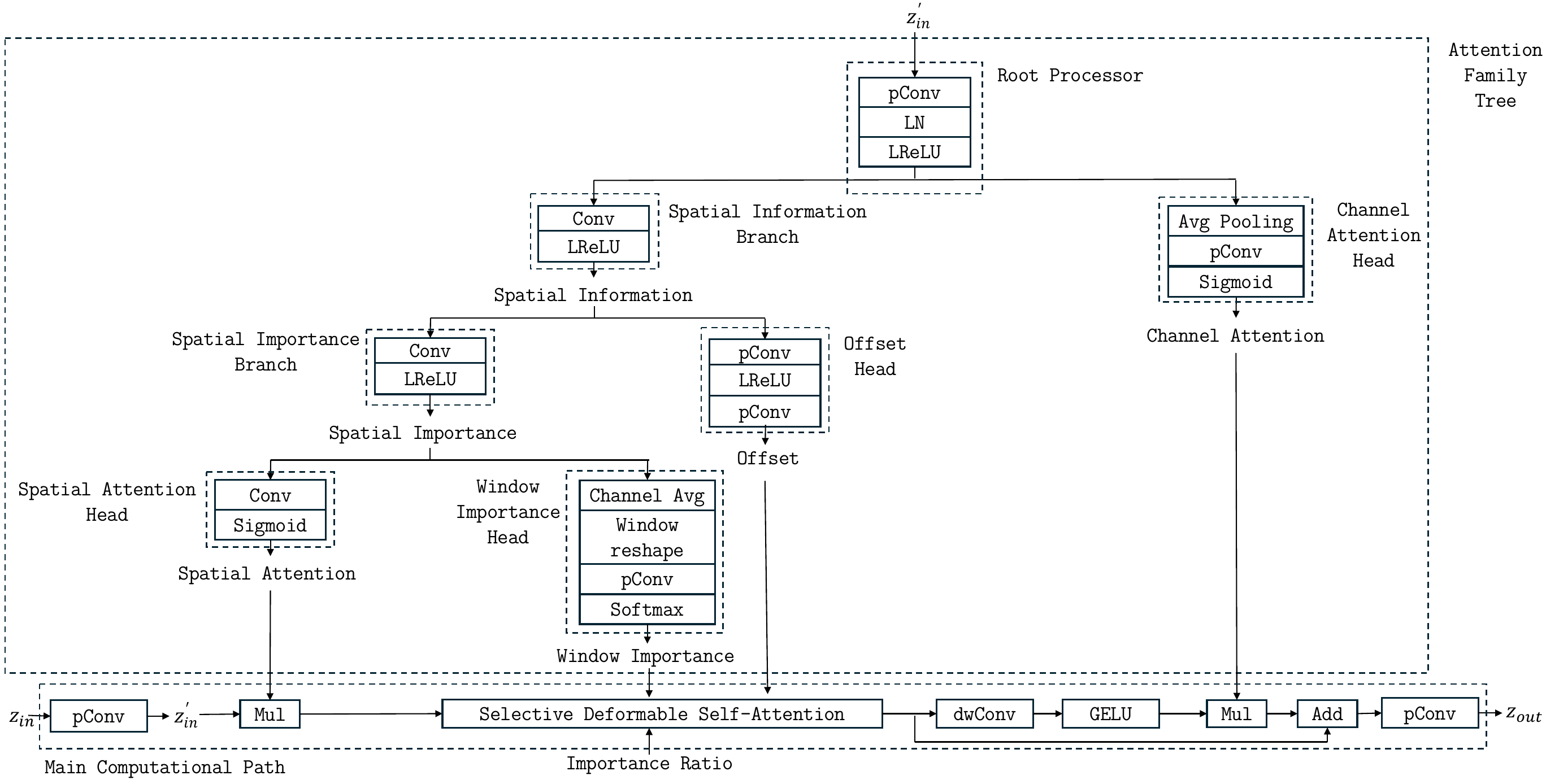}
    \caption{Dimension specialized operations of the proposed FA block.}\label{fig:dimension_specialized}
    \vspace{-0.15in}
\end{figure*}

In each stage, the feature importance-aware group (FA group), consisting of multiple feature importance-aware blocks (FA blocks), performs the core encoding and decoding processes of FAJSCC. To enhance communication accuracy while reducing computational resources, our FA block integrates axis-dimension specialized computation and selective deformable self-attention within an attention family tree. Figure~\ref{fig:dimension_specialized} depicts the overall procedure of the proposed dimension-specialized operations. The axis-dimension specialized computation refines features separately along the channel and spatial axes, reducing computational cost while representing features effectively. It corresponds to the main computational path except for the selective deformable self-attention module. The selective deformable self-attention applies self-attention only to the most important features by adaptively adjusting attention regions according to the input feature relations and the selected windows. Furthermore, the proposed attention family tree provides the attention features required for both axis-dimension specialized computation and selective deformable self-attention. This is achieved with highly efficient computation by eliminating redundant calculations based on output feature relationships. The following subsections present our proposed methods in detail.

\subsection{Axis-Dimension Specialized Computation}
\label{sec:axis computation}
The standard convolution operation in deep learning layers processes features simultaneously along the height, width, and channel axes. While this joint processing is effective for learning correlations across all dimensions, it incurs a high computational complexity of $\mathcal{O}(H_{\out}W_{\out}C_{\out}C_{\inn}s^2)$. Here, $H_{\out},W_{\out},C_{\out},C_{\inn}$ denote the output feature’s height, width, channels, and input feature's channels, respectively, and $s$ represents the kernel size. This complexity can be reduced without sacrificing representational ability by decomposing the operation into two axis-specific convolutions: depthwise and pointwise. Depthwise convolution (dwConv) processes the spatial dimensions (height and width) independently within each channel, whereas pointwise convolution (pConv) operates only along the channel axis. Together, they capture spatial and channel correlations with complexities of $\mathcal{O}(H_{\out}W_{\out}C_{\out}s^2)$ and $\mathcal{O}(H_{\out}W_{\out}C_{\out}C_{\inn})$, respectively, thereby reducing overall computational cost while achieving strong feature processing ability.

In addition to employing depthwise and pointwise convolutions, we apply a spatial attention operation (multiply input feature with spatial attention $\mathbf{z}_{sa} \in [0,1]^{H_{\inn} \times W_{\inn} \times 1}$) before the depthwise convolution and a channel attention operation (multiply input feature with channel attention $\mathbf{z}_{ca} \in [0,1]^{1 \times 1 \times C_{\inn}}$) before the pointwise convolution. Spatial attention emphasizes important features along the spatial axes, while channel attention emphasizes important features along the channel axes~\cite{woo2018cbam}. These attentions enhance the depthwise and pointwise convolutions by informing which features are important. Moreover, these lightweight attention operations incur only $\mathcal{O}(H_{\inn}W_{\inn}C_{\inn})$ complexity, imposing minimal computational overhead. Using depthwise and pointwise convolutions has also been adopted in recent deepJSCC research~\cite{yu2025novel}. However, the application of lightweight attentions to enhance these operations is a novel idea, and its effectiveness is further validated through the ablation study in Section~\ref{sec:ablation}, in comparison with using depthwise and pointwise convolutions alone.

\subsection{Selective Deformable Self-Attention}
\label{sec:selective attention}
Although the above axis-dimension specialized operations effectively refine features along each axis, relying solely on them may overlook relations among complex features. To efficiently capture these relations, we propose the deformable self-attention mechanism. The detailed self-attention operation for $\mathbf{Z}_{\inn} \in \mathbb{R}^{H_{\inn}W_{\inn} \times C_{\inn}}$ is
\begin{align}
\mathbf{Z}_{\out} = \mathrm{softmax}\left(\frac{\mathbf{Q}\mathbf{K}^\intercal}{\sqrt{C_{\inn}}}\right)\mathbf{V},
\end{align}
where $\mathbf{Q}, \mathbf{K}, \mathbf{V} \in \mathbb{R}^{H_{\inn}W_{\inn} \times C_{\inn}}$ are the query, key, and value matrices obtained from $\mathbf{Z}_{\inn}$ through learnable linear projections, respectively. This full self-attention has been shown to be highly effective in semantic communication~\cite{niu2025multimodal,zhang2024unified}, as it captures dependencies among complex features through feature similarity measurement with $\mathbf{Q}\mathbf{K}^\intercal$. However, its computational complexity $\mathcal{O}\left(4 H_{\inn} W_{\inn} C_{\inn}^2 + 2 (H_{\inn} W_{\inn})^2 C_{\inn}\right)$ becomes prohibitively large as image resolution increases. To alleviate this computational burden, many semantic communication systems adopt window-based self-attention~\cite{zhang2025semantic,wu2024cddm,yang2024swinjscc,cheng2025tcc,wang2025task,zhang2025snr}. Since this method applies self-attention only within fixed local windows, the computational complexity is reduced to $\mathcal{O}\left(\frac{H_{\inn} W_{\inn}}{n} \left(4 n C_{\inn}^2+ 2 n^2 C_{\inn}\right)\right)$ where $n$ is the number of features in the given window and $n \ll H_{\inn} W_{\inn}$ in most cases. While conventional window self-attention enhances computational efficiency, this method has two major limitations: It cannot capture highly related features across different windows, and analyzing features of every window is computationally expensive. We solve these problems by proposing selective deformable self-attention.

\begin{figure}[t]
    \centering
    \includegraphics[width=0.9\linewidth]{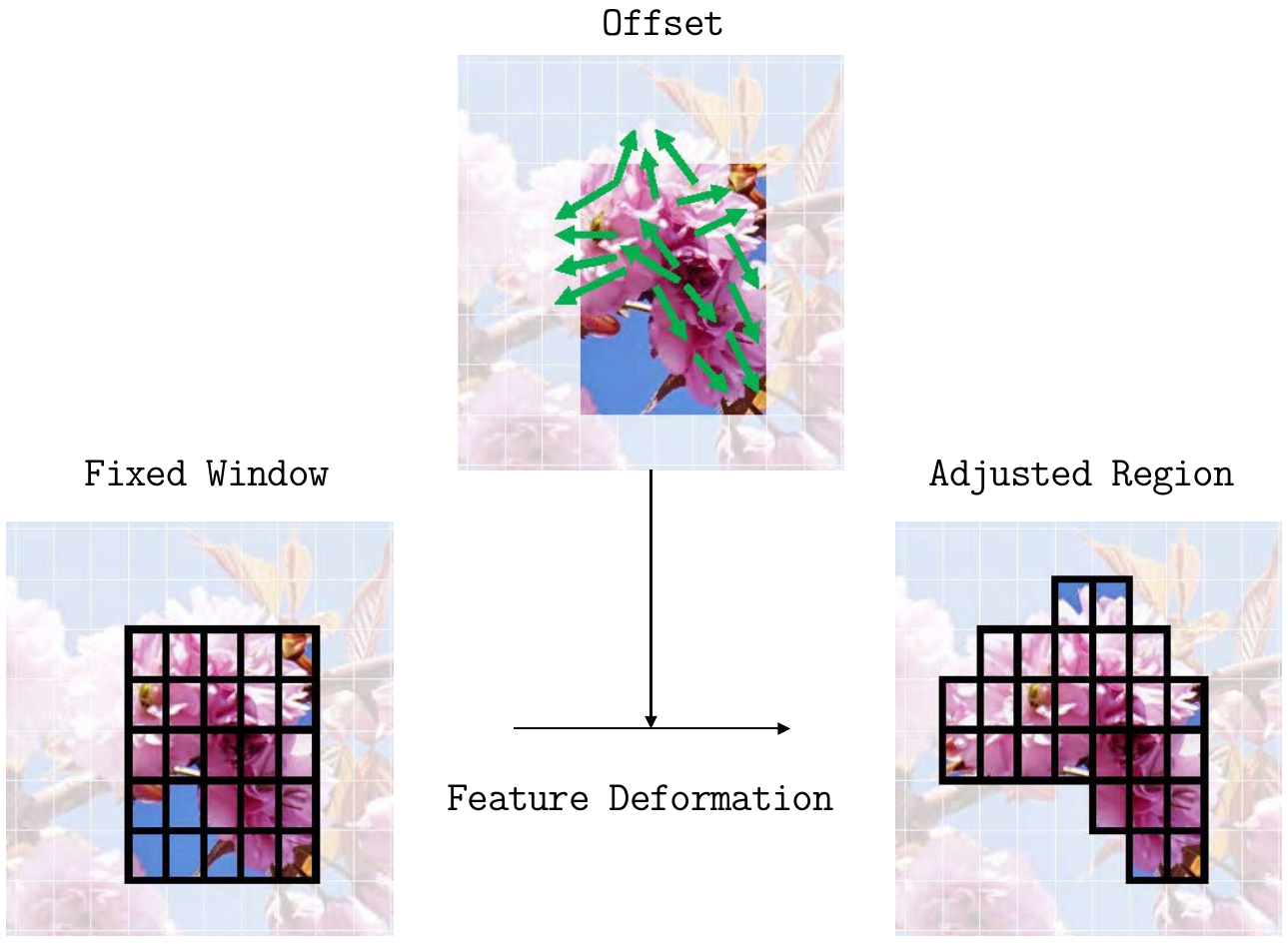}
    \caption{Visualization of feature deformation.}\label{fig:feature_deformation}
    \vspace{-0.15in}
\end{figure}

To capture strongly related features across different windows, we apply feature deformation before performing self-attention, which relocates related regions. The detailed process of feature deformation for the $i,j,c$-th height, width, and channel is as follows:
\begin{align}
& (u,v)=(i,j)+\mathbf{z}_{\offset}(i,j), \quad
i_0=\lfloor u\rfloor,\; i_1=i_0+1, \\
& j_0=\lfloor v\rfloor,\; j_1=j_0+1, \quad
\alpha = u-i_0, \;\; \beta = v-j_0, \\
& z_{\out}^c(i,j) =
\begin{bmatrix} 1-\alpha & \alpha \end{bmatrix}
\begin{bmatrix}
z_{\inn}^c(i_0,j_0) & z_{\inn}^c(i_0,j_1)\\
z_{\inn}^c(i_1,j_0) & z_{\inn}^c(i_1,j_1)
\end{bmatrix}
\begin{bmatrix} 1-\beta \\ \beta \end{bmatrix}.
\end{align}
Here, $(u,v)$ denotes the deformed sampling location predicted from the original coordinate $(i,j)$ and the predicted offset $\mathbf{z}_{\offset} \in \mathbb{R}^{H_{\inn} \times W_{\inn} \times 2}$ from the input feature relations. The four neighboring integer grid points $i_0,i_1,j_0,j_1$ represent the surrounding pixels (or features) that are closest to $(u,v)$. The interpolation weights $\alpha$ and $\beta$ correspond to the fractional distances between $(u,v)$ and the integer coordinates along the height and width axes, respectively.\footnote{The interpolation weights $\alpha$ and $\beta$ enable gradient flow into $\mathbf{z}_{\offset}$.} The equation for $z_{\out}^c(i,j)$ performs bilinear interpolation: When $(u,v)$ coincides with an integer grid point, it directly samples the corresponding feature value, whereas when $(u,v)$ lies between grid points, it computes a weighted combination of the four surrounding features, thereby interpolating the feature map at the non-integer position. This formulation ensures that both cases (direct alignment with existing feature coordinates and interpolation at in-between positions) are handled within a single equation, allowing subsequent self-attention to adaptively operate on adjusted regions instead of being limited to fixed grids. This effect is illustrated in Figure~\ref{fig:feature_deformation}.\footnote{Note that the figure (and subsequent figures as well) illustrates the deformation in the image domain to ease understanding, but the actual deformation operates in the feature domain.} The thick black borders in this figure indicate the features selected for self-attention. Applying self-attention within fixed-window areas may overlook feature correlations beyond them. In contrast, feature deformation enables self-attention to operate on adaptively adjusted regions, thereby capturing correlations that go beyond the fixed windows. We call this method deformable (window) self-attention.

The idea of feature deformation originates from deformable convolution~\cite{dai2017deformable,xiong2024efficient}, which has also been successfully applied in recent semantic communication systems~\cite{han2025scsc}. In addition, feature deformation for full self-attention has been explored in computer vision tasks~\cite{xia2022vision}, where the entire input feature space is deformed. However, such global deformation is not suitable for the selective enhancement framework that we introduce in the next paragraph, because it operates over the entire feature set and therefore cannot support selective refinement. In contrast, our method confines the deformation to features within a given window, enabling more targeted and efficient enhancement. Furthermore, to the best of our knowledge, feature deformation has not previously been applied to self-attention for constructing a novel deepJSCC architecture.

Deformable self-attention effectively enhances the capability of standard self-attention. Building on this, we propose to applying deformable self-attention selectively to improve computational efficiency. Previous studies in neural source coding have shown that not all features carry the same amount of information: Some features are more informative (e.g., main objects), while others are less informative (e.g., background regions)~\cite{cheng2020learned}. Moreover, it has been demonstrated that simple images do not require high computational cost to achieve high performance in super-resolution and image compression tasks~\cite{kong2021classsr,wang2024camixersr,hang2025rate}. Motivated by these observations, we apply deformable self-attention only to features in the most important windows, identified based on their contribution to image transmission performance. By restricting self-attention to these windows, the computational cost is significantly reduced. This selective enhancement strategy greatly reduces computational cost compared with applying deformable self-attention across all windows. A portion of the saved computation is then reinvested to increase intermediate feature channel sizes, leading to richer representations. Although this reinvestment slightly raises the complexity, the increase is much smaller than the computational savings, so the overall cost remains significantly lower than applying self-attention to features in all windows. As a result, FAJSCC achieves higher transmission performance by enhancing important feature processing while still maintaining reduced overall complexity.

\begin{figure*}[t]
    \centering
    \includegraphics[width=1.0\linewidth]{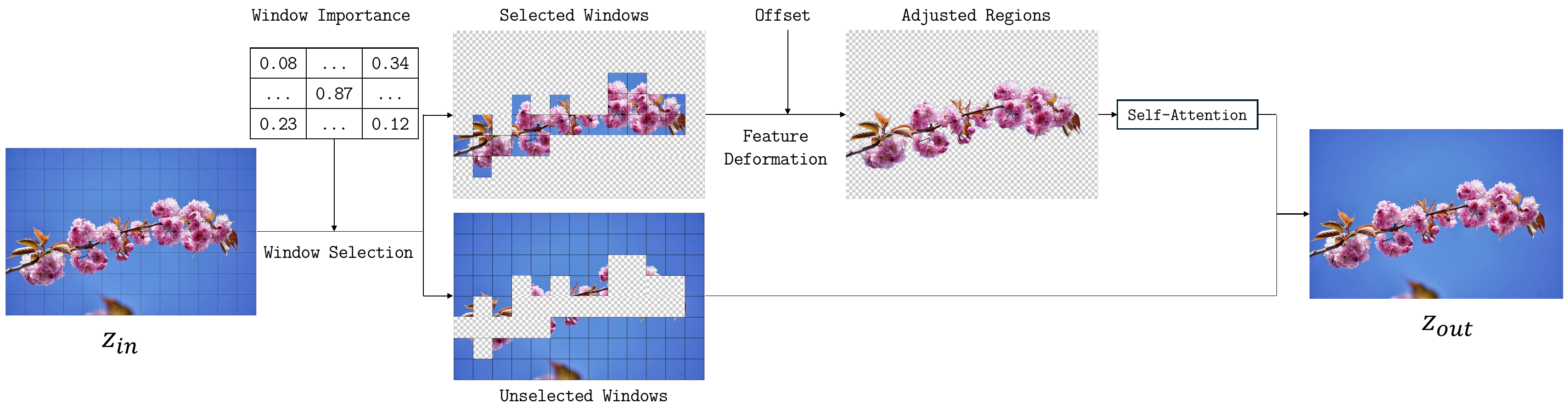}
    \caption{Procedures of selective deformable self-attention.}\label{fig:selective_attention}
    \vspace{-0.15in}
\end{figure*}

Figure~\ref{fig:selective_attention} gives the detailed procedures of our proposed selective deformable self-attention. Firstly, based on the window importance $\mathbf{z}_{\imp} \in [0,1]^{\frac{H_{\inn}}{\sqrt{n}} \times \frac{W_{\inn}}{\sqrt{n}}}$, windows are selected whether to be enhanced via deformable self-attention or not. The number of selected windows is determined by the importance ratio $\gamma$. When $\gamma = 1$, features in all windows are processed via deformable self-attention, whereas $\gamma = 0$ means none are processed. For $0<\gamma<1$, the features in the top $\gamma$ fraction windows, ranked by their measured importance, are selected and processed through deformable self-attention, which results in $\mathcal{O}\left(\frac{\gamma H_{\inn} W_{\inn}}{n} \left(4 n C_{\inn}^2+ 2 n^2 C_{\inn}\right)\right)$ computational complexity. Secondly, the feature deformation process of Figure~\ref{fig:feature_deformation} is applied to the selected windows to adjust self-attention regions based on the input feature relations. Lastly, self-attention is applied to the selected adjusted areas, and the original input feature shape is restored with processed selected features and unselected features. During the restoration phase, the deformed features used in selective self-attention are mapped back to their original spatial positions by applying the inverse transformation of the feature deformation process.

Unlike previous lightweight deepJSCC methods that reduce or remove features~\cite{jia2023lightweight,niu2025multimodal,zhang2024unified}, which often degrade performance, FAJSCC selectively enlarges and refines important features, achieving higher performance while reducing computational cost. Moreover, by adjusting the importance ratio, FAJSCC decouples computational complexity from the model architecture, allowing first independent evaluation of the impact of encoder and decoder complexity on performance. The prior resource-adjustable methods cannot achieve this, as the transceiver's computational complexity adjustments are constrained by the counterpart communication side~\cite{niu2025multimodal,zhang2024unified,raha2025dd}. Experiments varying encoder and decoder resources independently reveal, for the first time in the deepJSCC literature, that understanding the meaning of noisy features in the decoder demands the greatest computational cost. Detailed results are shown in Section~\ref{sec:main}.

\subsection{Attention Family Tree}

The proposed axis-dimension specialized operation and the selective deformable self-attention are both highly efficient and impose minimal computational overhead. However, we observe that na\"{i}vely computing the attention features, offsets, and window importance from scratch negates the efficiency gains of our methods; thus, we propose a new attention family tree that jointly extracts attention features and obtains offset and feature importance at once by eliminating overlapping calculations based on relations among output features to minimize computational resource overhead. The attention family tree structure is shown in Figure~\ref{fig:dimension_specialized}.

Note that the offset used in the deformation is obtained from the spatial-wise feature information, and spatial attention is obtained from the spatial-wise feature importance. The spatial-wise feature importance can be obtained from the spatial-wise feature information; thus, reusing the spatial-wise feature information to get spatial-wise feature importance can eliminate overlapping pre-processing calculations. Adding to this, spatial-wise feature importance can be directly used to identify which window is important, as well as obtain the spatial attention feature. Thus, the attention, offset, and window importance can all be obtained efficiently without a large computational burden. In ablation studies in Section~\ref{sec:ablation}, we show that our attention family tree shows better or similar performance while reducing computational burden compared to obtaining these attention features separately. Moreover, we also show that the efficiency gain comes from appropriately leveraging the relations of offset, attention feature, and window importance.

\subsection{Training and Loss Function Design}
\label{sec:training method}
Unlike in the test phase where simply selecting the top $\gamma$ fraction of windows suffices for selective deformable self-attention, such discrete selection prevents backpropagation during training. To address this issue, we employ Gumbel-Softmax sampling~\cite{jang2017categorical} during training. This approach replaces a non-differentiable selection process with a differentiable sampling process, enabling proper gradient flow.  More specifically, we sample important windows with probabilities proportional to window importance $\mathbf{z}_{\imp}$. By defining the probability that $(i,j)$-th coordinate window is sampled as important window $\pi^{1}(i,j):=z_{\imp}(i,j)$, and the probability that $(i,j)$-th coordinate window is not sampled as important window $\pi^{2}(i,j):=1-z_{\imp}(i,j)$, the soft sampling result $\mathbf{z}_{\soft}$ and the hard sampling result $\mathbf{z}_{\hard}$ are as follows:

\begin{align}
z^{l}_{\soft}(i,j) &= \frac{\exp{ \left(\log\pi^{l}(i,j)+g^{l}(i,j) \right)/\tau}}{\sum_{m=1}^2 \exp{ \left(\log\pi^{m}(i,j)+g^{m}(i,j) \right)/\tau}},\\
\mathbf{z}_{\hard}(i,j) &= \onehot \left(\argmax_{l} \left[\log \pi^{l}(i,j) +g^{l}(i,j) \right]    \right).
\end{align}
Here, $g^{l}(i,j)$ is i.i.d.~sample from $\text{Gumbel}(0,1)$,\footnote{This sampling can be done with $t^{l}(i,j)\sim$Uniform$(0,1)$, and $g^l(i,j)=- \log ( -\log ( t^{l}(i,j)))$.} $\tau$ is a hyperparameter, and $\onehot(\cdot)$ is a one-hot vector whose entry is $1$ at the selected index and $0$ elsewhere. 

The hard sampling result $\mathbf{z}_{\hard}$ follows the exact true distribution with $\mathbf{\pi}^{1},\mathbf{\pi}^{2}$ probabilities, but this prevents backpropagation. In contrast, the soft sampling result $\mathbf{z}_{\soft}$ enables backpropagation, but it deviates to some extent from the true distribution, except when $\tau=0$, i.e., $\mathbf{z}_{\soft}=\mathbf{z}_{\hard}$. To leverage the advantages of $\mathbf{z}_{\soft},\mathbf{z}_{\hard}$ both, we use the straight-through important window sampling result $\mathbf{z}_{\ST}$ as follows:\footnote{Straight-through is a technique for approximating gradients of non-differentiable functions to enable backward propagation.}
\begin{align}
\mathbf{z}_{\ST} = \mathbf{z}^{1}_{\hard}+\mathbf{z}^{1}_{\soft}-\stopg(\mathbf{z}^{1}_{\soft}).
\end{align}
Here, $\stopg(\cdot)$ indicates the stop-gradient operation, which behaves as the identity function in the forward propagation and outputs zero gradient in the backward propagation. The use of $\mathbf{z}_{\ST}$ enables discrete sampling during the forward process while preserving gradient flow at the backward process during training. In selective deformable self-attention, we use $\mathbf{z}_{\ST}$ to model the discrete sampling process and enable gradient flow as follows:
\begin{align}
\mathbf{z}_{\out} = \selfa( \defo ( \mathbf{z}_{\ST} \odot_{\textrm{w}} \mathbf{z}_{\inn} ) ).
\end{align}
Here, $\odot_{\textrm{w}}$ denotes window-wise multiplication that multiplies $z_{\ST}(i,j)$ to the features in the $(i,j)$-th window of $\mathbf{z}_{\inn}$. Due to gradient flow via $\mathbf{z}^{1}_{\soft}$ term of $\mathbf{z}_{\ST}$, the window importance head in Figure~\ref{fig:dimension_specialized} can be optimized with respect to a given loss function during training. Recall that this Gumbel-Softmax method is only used in the training phase. In the test phase, since gradient flow is not required, we do not use $\mathbf{z}_{\ST}$ and simply select important windows in $\mathbf{z}_{\inn}$ based on $\mathbf{z}_{\imp}$ as detailed in Section~\ref{sec:selective attention}.

To encourage the window importance head to assign a high score to $z_{\imp}(i,j)$ when the $(i,j)$-th window is important and a low score otherwise, we introduce a penalty term that discourages high scores for less important windows. Let $\{ \mathbf{x}^{i} \}_{i=1}^{N}$ denote an image set of batch size $N$, where $\mathbf{x}^{i}$ represents the $i$-th image. To obtain the desired importance score, we train our FAJSCC using the following feature importance-aware loss function, denoted as $\mathcal{L}_{\FA}$:
\begin{align}
    \mathcal{L}_{\FA} &:= \frac{1}{N} \sum_{i=1}^{N} \mathcal{L}_{\distortion}(\mathbf{x}^{i},\hat{\mathbf{x}}^{i}) + \eta \norm{\gamma_{\tr} - \frac{( \bar{\mathbf{z}}^{e}_{\ST} + \bar{\mathbf{z}}^{d}_{\ST} ) }{2}}_{2}^{2},\label{eq:FAJSCC loss}\\
    \bar{\mathbf{z}}^{e}_{\ST} &= \frac{1}{N} \sum_{i=1}^{N}     \bar{\mathbf{z}}^{e,i}_{\ST},~~\bar{\mathbf{z}}^{d}_{\ST} = \frac{1}{N} \sum_{i=1}^{N} \bar{\mathbf{z}}^{d,i}_{\ST}, ~~0<\gamma_{\tr}<1,
\end{align}
where $\eta$ is the sampling weight, $\gamma_{\tr}$ is the target average importance ratio in the training phase, and $\bar{\mathbf{z}}^{e,i}_{\ST}$ and $\bar{\mathbf{z}}^{d,i}_{\ST}$ represent the mean of $\mathbf{z}_{\ST}$ values for the $i$-th image in the encoder and decoder, respectively. The first term $\mathcal{L}_{\distortion}(\mathbf{x},\hat{\mathbf{x}})$ is an image distortion loss between original image $\mathbf{x}$ and reconstructed image $\hat{\mathbf{x}}$, as will be detailed in Section~\ref{sec:experiment_setting}. To minimize both $\mathcal{L}_{\distortion}(\mathbf{x},\hat{\mathbf{x}})$ and $\norm{\gamma_{\tr} - \frac{( \bar{\mathbf{z}}^{e}_{\ST} + \bar{\mathbf{z}}^{d}_{\ST} ) }{2}}_{2}^{2}$ in $\mathcal{L}_{\FA}$, the window importance should be high for windows that contribute significantly to performance improvement and low for those with limited impact when processed via deformable self-attention. 

As a result, minimizing $\mathcal{L}_{\FA}$ with Gumbel-Softmax sampling enables FAJSCC to efficiently reduce the distortion between $\mathbf{x}$ and $\hat{\mathbf{x}}$ by appropriately sampling important windows. As the Gumbel-Softmax method samples important windows with probabilities proportional to their window importance, $\bar{\mathbf{z}}^{e,i}_{\ST}$ and $\bar{\mathbf{z}}^{d,i}_{\ST}$ fluctuate randomly for each $i$-th image transmission while keeping the batch-wise averages $\bar{\mathbf{z}}^{e}_{\ST}$ and $\bar{\mathbf{z}}^{d}_{\ST}$ stable during training. This enables FAJSCC to adapt to various computational budgets while ensuring stable updates during training. Another benefit of using Gumbel-Softmax is that it allows independent control of the importance ratios for the encoder and decoder during the test phase. This plays a crucial role in the experiment that measures the impact of computational complexity, which will be discussed later.

Incorporating Gumbel–Softmax sampling and the regularization term in Eq.~\ref{eq:FAJSCC loss} (3.62 ms per image) introduces only a marginal training overhead, compared to training with $\mathcal{L}_{\distortion}$ alone using random sampling without the window-importance head of Figure~\ref{fig:dimension_specialized} (3.43 ms per image). In addition, the convergence behavior obtained with Eq.~\ref{eq:FAJSCC loss} exhibits a similar trend to that of previous deepJSCC methods~\cite{bourtsoulatze2019deep,zhang2023predictive,yang2024swinjscc,yu2025novel}. Furthermore, setting $\gamma_{\tr} \in [0.4, 0.6]$ does not lead to noticeable performance differences. However, excessively small or large values of $\gamma_{\tr}$ should be avoided. A too-small $\gamma_{\tr}$ leads to insufficient training of the selective deformable self-attention block, whereas a too-large $\gamma_{\tr}$ results in inadequate training of the window-importance head, i.e., all features are always selected as important. Note that $\gamma_{\tr}$ is a training hyperparameter, similar in role to the learning rate.

\section{Experiment}
\label{sec:experiment}

\subsection{Experimental Setting}
\label{sec:experiment_setting}

\begin{table}
\centering
\caption{Architecture Summary Table for FAJSCC.}
\label{tab:arc_summary}
\begin{tabular}{|l||l|}
\hline
The number of stages ($L$)                & $4$ \\ \hline
The number of FA blocks per stage ($n_i$) & $[2,2,2,2]$ \\ \hline
Feature channel sizes ($C_i$)     & $[40,60,80,260]$ \\ \hline
Window sizes                      & $[8,8,8,8]$ \\ \hline
The number of attention heads             & $[1,1,1,1]$ \\ \hline
\end{tabular}
\end{table}

\noindent \textbf{Baselines and Architectures:} We implement two versions of FAJSCC: The first is as illustrated in Figure~\ref{fig:deepJSCCarchitecture} with details in Section~\ref{sec:FAJSCC framework}. The other is without the selective deformable self-attention proposed in Section~\ref{sec:selective attention}, i.e., unused offset and window importance heads in the attention family tree, and the selective deformable self-attention module in the main computational path are removed. We call the latter version light-attention deepJSCC (LAJSCC). LAJSCC uses only the axis-dimension specialized computation proposed in Section~\ref{sec:axis computation} to verify its computational efficiency. The other deepJSCC baselines range from the original deepJSCC~\cite{bourtsoulatze2019deep} to the recent SwinJSCC~\cite{yang2024swinjscc} and LICRFJSCC~\cite{yu2025novel}. 

To differentiate between these baselines, we refer to the initial deepJSCC~\cite{bourtsoulatze2019deep} based on the convolution blocks and the deepJSCC based on residual convolution blocks~\cite{zhang2023predictive} as ConvJSCC and ResJSCC, respectively. In contrast to SwinJSCC~\cite{yang2024swinjscc} that applies self-attention to all features, LICRFJSCC~\cite{yu2025novel} applies self-attention only to a fixed subset of feature channels. The scaling settings for ConvJSCC and ResJSCC follow those in~\cite{bourtsoulatze2019deep}, while the scaling configuration for SwinJSCC is determined via an exhaustive search to achieve the best possible performance under comparable GFLOPs constraints. LICRFJSCC, LAJSCC, and FAJSCC scale settings are based on SwinJSCC. The architecture summary of FAJSCC is provided in Table~\ref{tab:arc_summary}. In addition, the detailed implementation code is available at github.com/hansung-choi/FAJSCCv2.

\medskip
\noindent \textbf{Datasets and Training} To train models, we use the DIV2K training dataset~\cite{agustsson2017ntire}, which consists of $800$ high-resolution 2K images. To maintain consistent image resolution within each batch and prevent out-of-memory errors, we randomly crop images to $128 \times 128$ during training, with a batch size of $32$. The image distortion loss $\mathcal{L}_{\distortion}$ is set to $\MSE$ and $1-\SSIM$ when evaluating models for PSNR and SSIM, respectively. Our FAJSCC is trained to minimize $\mathcal{L}_{\FA}$ in Eq.~\eqref{eq:FAJSCC loss} while others are trained to minimize $\mathcal{L}_{\distortion}$. The hyperparameter settings for $\mathcal{L}_{\FA}$ are $\eta=1.0$, $\gamma_{\tr}=0.5$ , and $\tau=1$. Note that $\gamma_{\tr}$ is the target average importance ratio in the training phase; the importance ratio during the test phase could be different, e.g., Figure~\ref{fig:resource control}. All models are trained using Adam optimizer with a learning rate $0.0001$ for $200$ epochs. Experiments were conducted on a system equipped with dual Intel Xeon E5-2650 v3 CPUs (10 cores in total) and NVIDIA Tesla P100 GPU with 16 GB memory.

We consider four CPP values $[\frac{1}{12},\frac{1}{16},\frac{1}{24},\frac{1}{32}]$ and four SNR values $[1,4,7,10]$dB under AWGN and fast Rayleigh fading channels. Except for SNR robustness experiments, the training SNR is the same as the test SNR. We evaluated deepJSCC models trained with $5$ different random seeds. For tables, values are reported as $\mean \pm \std$, while entries with $0.00$ $\std$ (e.g., model storage size) are omitted. For efficiency comparison plots, $\std$ values are described with vertical lines. For other plots, we omitted the standard deviation to maintain visual clarity and avoid excessive and repeated clutter.

Test results are obtained from the DIV2K validation dataset~\cite{agustsson2017ntire} and the Kodak dataset~\cite{bworld}. The heights and widths of the DIV2K validation images range from $1068$ to $2040$ pixels. To preserve the original spatial characteristics as closely as possible, while aligning with the SwinJSCC's input resolution constraint, each image is center-cropped to the nearest multiple of $128$. The Kodak dataset consists of images with resolutions of $512 \times 768$ or $768 \times 512$. Owing to the variability in image resolutions, these datasets enable a realistic evaluation of FAJSCC and LAJSCC in practical applications, such as monitoring scenarios.

\begin{figure*}
\centering
\includegraphics[width=0.8\linewidth]{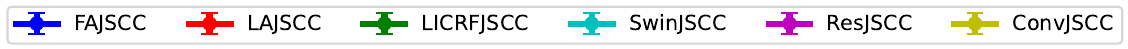}\\
\begin{multicols}{2}
\centering
\includegraphics[width=0.95\linewidth]{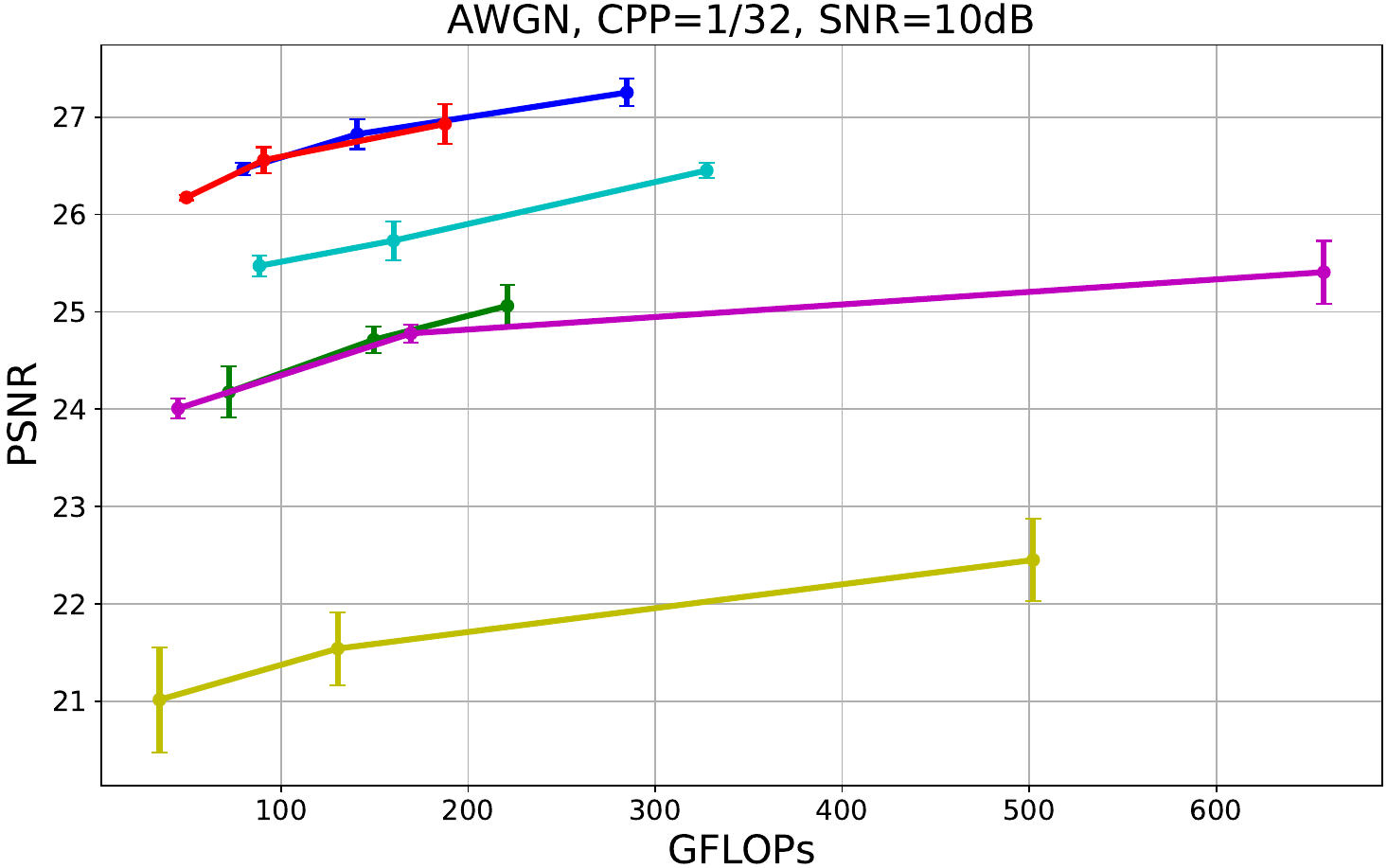}\\

\includegraphics[width=0.95\linewidth]{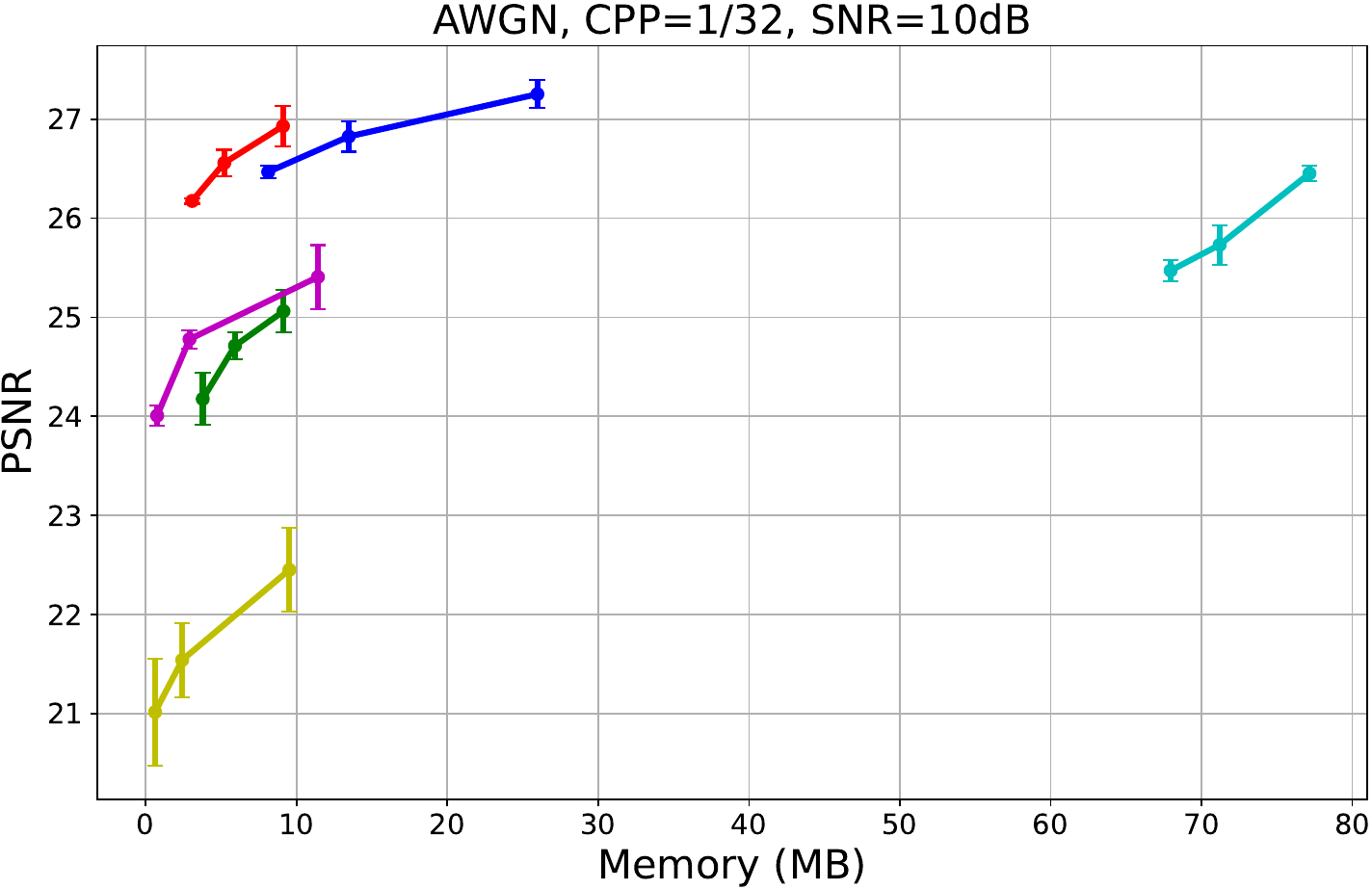}\\

\end{multicols}

\vspace{-0.15in}
\caption{Efficiency comparison for various model sizes with respect to computational burden (GFLOPs) and model storage size (MB) for DIV$2$K dataset.}\label{fig:efficiency comparison}
\vspace{-0.1in}
\end{figure*}

\begin{table*}
\centering
\caption{Comparison for deployment costs of deepJSCCs under AWGN channel, SNR=$10\ \mathrm{dB}$, and CPP=$1/32$ for Kodak dataset.}
\label{tab:Kodak_Deployment_Cost}
\begin{tabular}{|l||l|l|l|l|l|}
\hline
Metric               & PSNR (dB)                 & Latency (ms/image)         & GFLOPs  & Peak Memory (MB) & Model Storage Size (MB) \\ \hline\hline
ConvJSCC             & $20.71${\tiny $\pm 0.31$} & $4.37${\tiny $\pm 0.67$}   & $4.36$  & $46.78$  & $0.65$   \\ \hline
ResJSCC              & $23.45${\tiny $\pm 0.17$} & $13.48${\tiny $\pm 0.74$}  & $5.60$  & $59.62$  & $0.79$   \\ \hline
small SwinJSCC       & $25.15${\tiny $\pm 0.07$} & $96.64${\tiny $\pm 1.52$}  & $11.04$ & $441.16$ & $68.79$  \\ \hline
LICRFJSCC            & $24.84${\tiny $\pm 0.31$} & $45.45${\tiny $\pm 0.88$}  & $9.00$  & $311.16$ & $3.88$   \\ \hline
small LAJSCC (ours)  & $25.57${\tiny $\pm 0.04$} & $55.48${\tiny $\pm 1.42$}  & $6.16$  & $143.94$ & $3.08$   \\ \hline
small FAJSCC (ours)  & $25.77${\tiny $\pm 0.07$} & $100.25${\tiny $\pm 1.91$} & $9.96$  & $213.74$ & $8.12$   \\ \hline\hline
large ConvJSCC       & $21.91${\tiny $\pm 0.43$} & $8.39${\tiny $\pm 0.53$}   & $16.27$ & $77.37$  & $2.44$   \\ \hline
large ResJSCC        & $24.24${\tiny $\pm 0.13$} & $16.98${\tiny $\pm 0.72$}  & $21.16$ & $102.36$ & $2.90$    \\ \hline
SwinJSCC             & $25.40${\tiny $\pm 0.15$} & $99.15${\tiny $\pm 1.58$}  & $20.00$ & $493.38$ & $72.14$  \\ \hline
large LICRFJSCC      & $25.33${\tiny $\pm 0.20$} & $59.66${\tiny $\pm 0.63$}  & $18.68$ & $366.43$ & $5.92$   \\ \hline
LAJSCC (ours)        & $25.79${\tiny $\pm 0.12$} & $59.76${\tiny $\pm 0.76$}  & $11.33$ & $198.70$ & $5.31$  \\ \hline
FAJSCC (ours)        & $26.30${\tiny $\pm 0.12$} & $112.01${\tiny $\pm 2.29$} & $17.52$ & $303.86$ & $13.62$  \\ \hline\hline
huge ConvJSCC        & $21.45${\tiny $\pm 0.60$} & $22.58${\tiny $\pm 0.04$}  & $62.75$ & $145.20$ & $9.99$    \\ \hline
huge ResJSCC         & $24.17${\tiny $\pm 0.29$} & $39.64${\tiny $\pm 0.04$}  & $82.19$ & $197.68$ & $11.89$   \\ \hline
large SwinJSCC       & $25.77${\tiny $\pm 0.12$} & $115.49${\tiny $\pm 1.65$} & $40.93$ & $567.06$ & $78.00$  \\ \hline
huge LICRFJSCC       & $25.78${\tiny $\pm 0.27$} & $70.57${\tiny $\pm 0.61$}  & $27.61$ & $411.87$ & $9.13$    \\ \hline
large LAJSCC (ours)  & $26.39${\tiny $\pm 0.09$} & $70.50${\tiny $\pm 1.50$}  & $23.44$ & $287.98$ & $9.10$  \\ \hline
large FAJSCC (ours)  & $26.67${\tiny $\pm 0.09$} & $130.15${\tiny $\pm 2.56$} & $35.63$ & $441.68$ & $25.96$  \\ \hline
\end{tabular}
\end{table*}

\begin{table*}
\centering
\caption{Comparison for deployment costs of deepJSCCs under AWGN channel, SNR=$10\ \mathrm{dB}$, and CPP=$1/32$ for DIV$2$K dataset.}
\label{tab:DIV2K_Deployment_Cost}
\begin{tabular}{|l||l|l|l|l|l|}
\hline
Metric               & PSNR (dB)                 & Latency (ms/image) & GFLOPs & Peak Memory (MB) & Model Storage Size (MB) \\ \hline\hline
ConvJSCC             & $21.09${\tiny $\pm 0.43$} & $21.57${\tiny $\pm 0.25$}   & $34.91$  & $301.87$  & $0.65$   \\ \hline
ResJSCC              & $24.05${\tiny $\pm 0.10$} & $44.07${\tiny $\pm 0.36$}   & $44.86$  & $402.65$  & $0.79$   \\ \hline
small SwinJSCC       & $25.09${\tiny $\pm 0.11$} & $690.25${\tiny $\pm 2.54$}  & $88.33$  & $3017.08$ & $68.79$  \\ \hline
LICRFJSCC            & $24.52${\tiny $\pm 0.26$} & $251.84${\tiny $\pm 0.37$}  & $72.00$  & $2348.25$ & $3.88$    \\ \hline
small LAJSCC (ours)  & $26.21${\tiny $\pm 0.02$} & $169.91${\tiny $\pm 1.10$}  & $49.29$  & $1020.79$ & $3.08$    \\ \hline
small FAJSCC (ours)  & $26.39${\tiny $\pm 0.06$} & $352.488${\tiny $\pm 1.84$} & $79.66$  & $1519.75$ & $8.12$    \\ \hline\hline
large ConvJSCC       & $22.24${\tiny $\pm 0.37$} & $43.30${\tiny $\pm 0.46$}   & $130.23$ & $521.46$  & $2.44$     \\ \hline
large ResJSCC        & $24.73${\tiny $\pm 0.09$} & $84.66${\tiny $\pm 0.31$}   & $169.35$ & $714.45$  & $2.90$    \\ \hline
SwinJSCC             & $25.73${\tiny $\pm 0.19$} & $704.51${\tiny $\pm 2.88$}  & $160.04$ & $3347.44$ & $72.14$    \\ \hline
large LICRFJSCC      & $24.98${\tiny $\pm 0.14$} & $342.51${\tiny $\pm 0.48$}  & $149.46$ & $2784.53$ & $5.92$    \\ \hline
LAJSCC (ours)        & $26.55${\tiny $\pm 0.13$} & $228.25${\tiny $\pm 1.06$}  & $90.66$  & $1421.05$ & $5.31$    \\ \hline
FAJSCC (ours)        & $26.82${\tiny $\pm 0.15$} & $355.74${\tiny $\pm 1.73$}  & $140.16$ & $2122.69$ & $13.62$   \\ \hline\hline
huge ConvJSCC        & $22.60${\tiny $\pm 0.42$} & $130.48${\tiny $\pm 0.43$}  & $502.05$ & $968.80$  & $9.99$    \\ \hline
huge ResJSCC         & $24.99${\tiny $\pm 0.32$} & $225.33${\tiny $\pm 0.46$}  & $657.52$ & $1356.65$ & $11.89$   \\ \hline
large SwinJSCC       & $26.30${\tiny $\pm 0.18$} & $854.70${\tiny $\pm 2.56$}  & $327.51$ & $3900.57$ & $78.00$  \\ \hline
huge LICRFJSCC       & $25.33${\tiny $\pm 0.21$} & $428.48${\tiny $\pm 0.97$}  & $220.97$ & $3078.96$ & $9.13$    \\ \hline
large LAJSCC (ours)  & $27.09${\tiny $\pm 0.20$} & $347.64${\tiny $\pm 0.41$}  & $187.55$ & $2089.58$ & $9.10$  \\ \hline
large FAJSCC (ours)  & $27.32${\tiny $\pm 0.14$} & $522.92${\tiny $\pm 1.39$}  & $284.96$ & $3111.11$ & $25.96$  \\ \hline
\end{tabular}
\end{table*}

\subsection{Main Result}
\label{sec:main}

\medskip
\noindent \textbf{Architecture Efficiency:} 

To verify the performance gain of our model, we provide an efficiency comparison for various model sizes in Figure~\ref{fig:efficiency comparison}. To match the computational resource usage for comparison, we implement smaller and larger versions of SwinJSCC, LAJSCC, and FAJSCC by decreasing and increasing feature channel sizes by about $50\%$. In the case of ConvJSCC and ResJSCC, we increase the feature channel size by factors of $2$ and $4$ to construct larger models. 

Figure~\ref{fig:efficiency comparison} shows the tradeoff between computational block complexity (GFLOPs and memory usage) and PSNR performance for various JSCC schemes. Most existing works, such as ConvJSCC~\cite{bourtsoulatze2019deep}, ResJSCC~\cite{zhang2023predictive}, and SwinJSCC~\cite{yang2024swinjscc}, primarily aim to improve PSNR. These performance gains are achieved by employing heavier computational blocks, which consequently incur substantially higher computational burden (GFLOPs) and model storage size (MB). In particular, SwinJSCC demands approximately an order of magnitude more model storage size than ConvJSCC and ResJSCC, and imposes a significantly higher computational burden for the same feature scales due to its reliance on heavy multi-head self-attention mechanisms. This increased requirement gives a large burden to practical deployment, especially for IoT devices. To address this issue, recent works decrease computational burden at the expense of performance degradation. For example, LICRFJSCC~\cite{yu2025novel} applies advanced self-attention only to a fixed subset of feature channels without considering feature importance. Although this method reduces the computational burden, it also compromises performance by shrinking the feature representation space. 

Different from such previous directions, our method achieves performance improvement while reducing computational burden, i.e., going to the upper left in the left plot of Figure~\ref{fig:efficiency comparison}. Our FAJSCC is located in the far upper-left corner by applying computationally expensive self-attention only to the important features. Moreover, by disabling the selectively deformable self-attention block, the FAJSCC model becomes LAJSCC. In this figure, FAJSCC's PSNR vs. GFLOPs is similar to that of the LAJSCC. This verifies that FAJSCC can dynamically control computational resources without compromising the performance efficiency gain in PSNR vs. GFLOPs achieved by LAJSCC. 

\medskip
\noindent \textbf{Deployment Cost for Kodak dataset:} For our NVIDIA Tesla P100 GPU, the models show no computation and memory traffic bottleneck for $512 \times 768$ Kodak images. When peak memory is not high compared to the GPU memory bandwidth, the GPU computation follows the next structure.
\begin{equation*}
\mathrm{Memory} \rightarrow \mathrm{Compute} \rightarrow \mathrm{Compute} \rightarrow \mathrm{Store}
\end{equation*}
In this case, the overall latency is not mainly affected by the memory traffic. Moreover, if the required GFLOPs of each block's forward pass in the model are also not high compared to the GPU computation ability, the overall latency is also not mainly affected by the total GFLOPs. Table~\ref{tab:Kodak_Deployment_Cost} precisely shows this relation. For example, ResJSCC ($5.60$ GFLOPs) has only $28.44 \%$ larger GFLOPs than ConvJSCC ($4.36$ GFLOPs), but the latency is more than twice. Moreover, although GFLOPs and peak memory usage of SwinJSCC ($20.00$ GFLOPs, $493.38$ MB) are larger than our FAJSCC ($17.52$ GFLOPs, $303.86$ MB), SwinJSCC ($99.15$ ms/image) is faster than FAJSCC ($112.01$ ms/image) by about $12.97 \%$. 

When neither computational capacity nor memory bandwidth is a bottleneck, the overall latency is largely determined by the total number of layers in the model, as both computationally intensive and lightweight operations can be executed concurrently within the GPU’s processing capability. Table~\ref{tab:Kodak_Deployment_Cost} follows this case. Since this case does not give any burden to GPU, the deployment costs are not that important. However, as image resolution increases, deployment costs become more important, as shown in the next analysis.

\medskip
\noindent \textbf{Deployment Cost for DIV$2$K dataset:} For our NVIDIA Tesla P100 GPU, previous models using self-attention show a memory traffic bottleneck for high-resolution $2$K DIV$2$K images, e.g., $1536 \times 2048$ resolution. Within each window for window-based self-attention, the computation of $QK^{\top}$, followed by softmax normalization and the subsequent multiplication with $V$, requires the formation and storage of intermediate tensors. Each window independently generates attention score matrices and intermediate activations, resulting in repeated global memory reads and writes. This memory-intensive behavior significantly increases off-chip memory traffic. In this case, the GPU computation follows the next structure.
\begin{equation*}
\mathrm{Load} \rightarrow \mathrm{wait} \rightarrow \mathrm{Compute} \rightarrow \mathrm{Store} \rightarrow \mathrm{wait} \rightarrow \mathrm{Load}
\end{equation*}
These repeated loading and waiting lead to excessive latency. For example, in Table~\ref{tab:DIV2K_Deployment_Cost}, huge ConvJSCC model requires $502.05$ GFLOPs and small SwinJSCC only requires $88.33$ GFLOPs. However, huge ConvJSCC ($130.48$ ms/image) is much faster than small SwinJSCC ($690.25$ ms/image) due to the small memory traffic.

The primary advantage of FAJSCC over SwinJSCC arises from the reduction in the number of processed windows via the importance ratio $\gamma$. Only half of the windows are selected for computation in FAJSCC in Table~\ref{tab:DIV2K_Deployment_Cost}. Consequently, compared to SwinJSCC, memory traffic is reduced significantly, and then, FAJSCC ($355.74$ ms/image) achieves less than half the latency of SwinJSCC ($704.51$ ms/image).

\begin{figure*}
\centering
\includegraphics[width=0.8\linewidth]{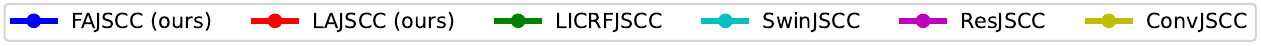}\\
\begin{multicols}{4}
\centering
\includegraphics[width=1.0\linewidth]{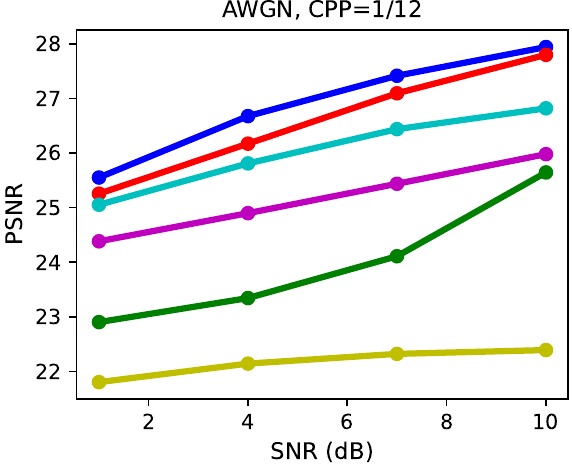}\\

\includegraphics[width=1.0\linewidth]{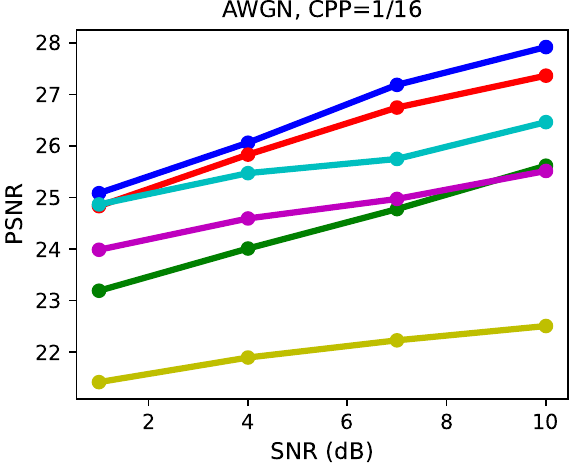}\\

\includegraphics[width=1.0\linewidth]{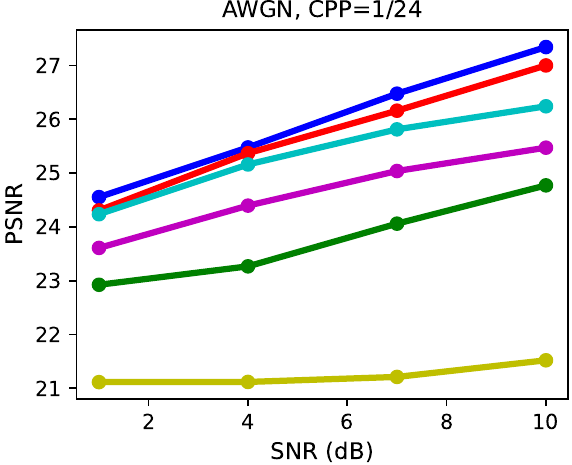}\\

\includegraphics[width=1.0\linewidth]{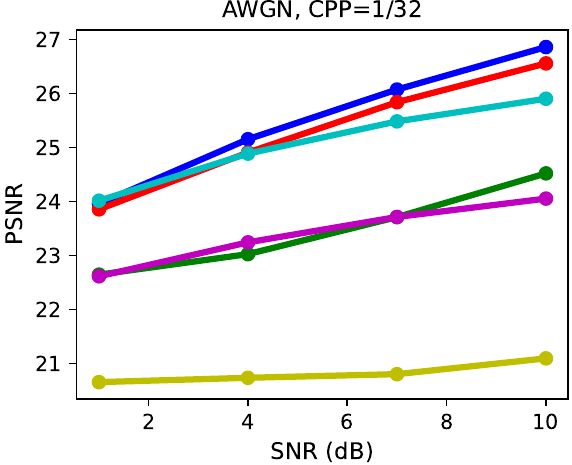}\\
\end{multicols}
\begin{multicols}{4}
\centering
\includegraphics[width=1.0\linewidth]{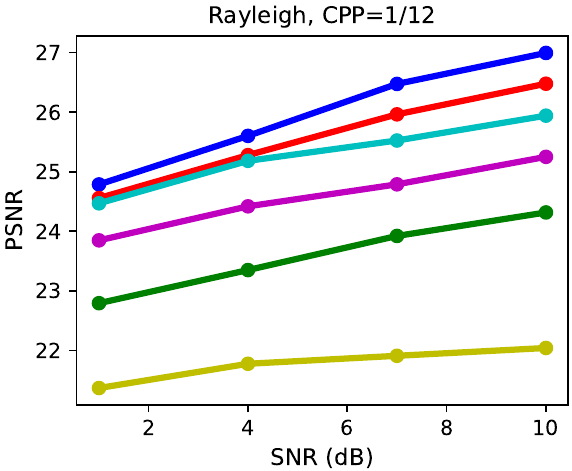}\\

\includegraphics[width=1.0\linewidth]{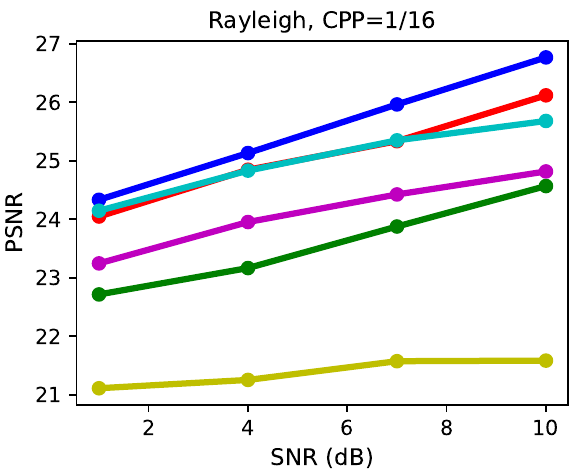}\\

\includegraphics[width=1.0\linewidth]{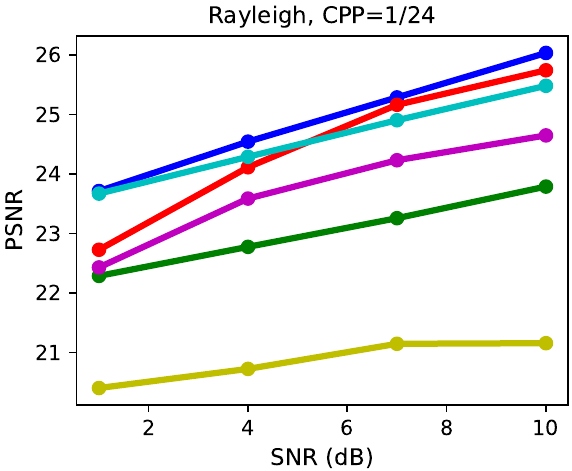}\\

\includegraphics[width=1.0\linewidth]{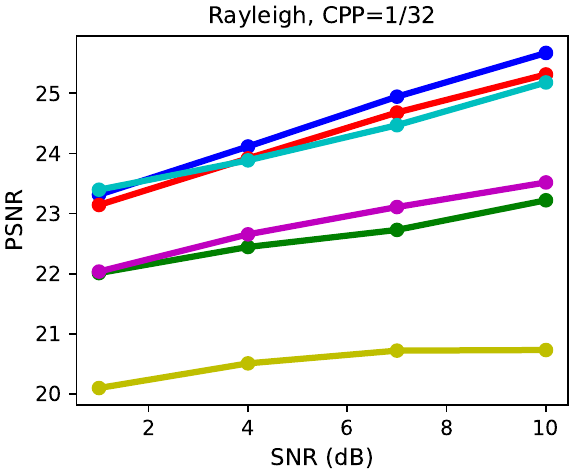}\\
\end{multicols}

\vspace{-0.15in}
\caption{PSNR results under different channel and CPP environments for DIV$2$K dataset.}\label{fig:psnr_result}
\vspace{-0.1in}
\end{figure*}

\begin{figure*}
\centering
\includegraphics[width=0.8\linewidth]{Figure3/plot_legend_main.pdf}\\
\begin{multicols}{4}
\centering
\includegraphics[width=1.0\linewidth]{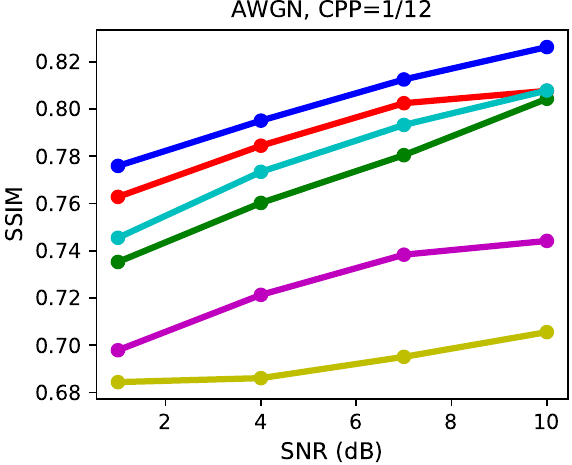}\\

\includegraphics[width=1.0\linewidth]{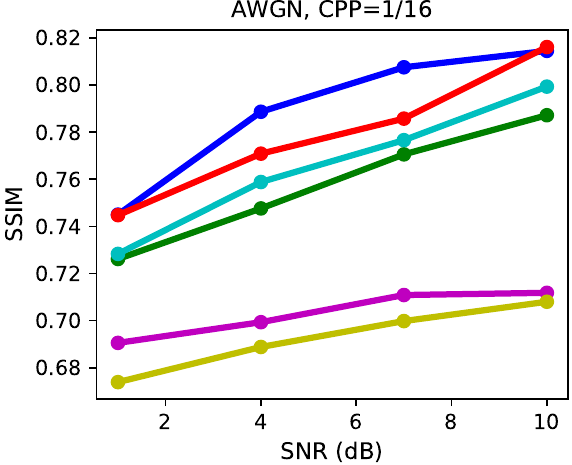}\\

\includegraphics[width=1.0\linewidth]{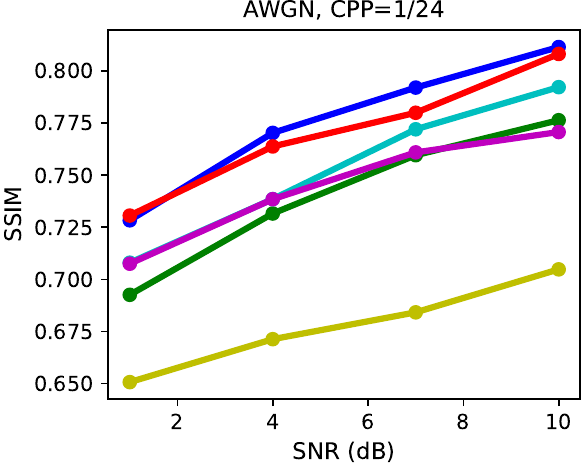}\\

\includegraphics[width=1.0\linewidth]{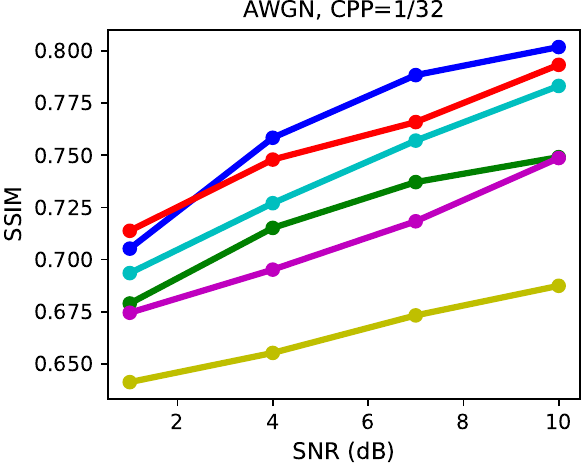}\\
\end{multicols}
\begin{multicols}{4}
\centering
\includegraphics[width=1.0\linewidth]{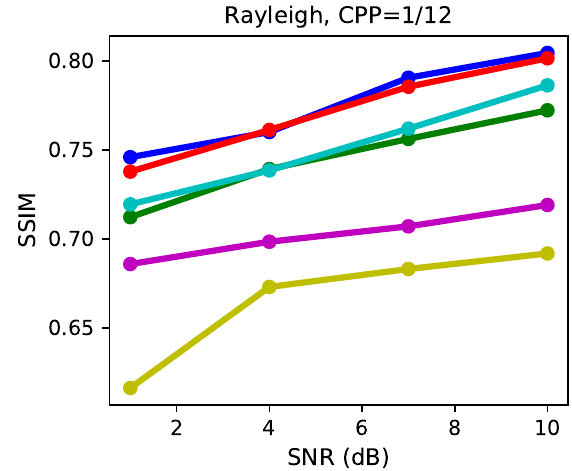}\\

\includegraphics[width=1.0\linewidth]{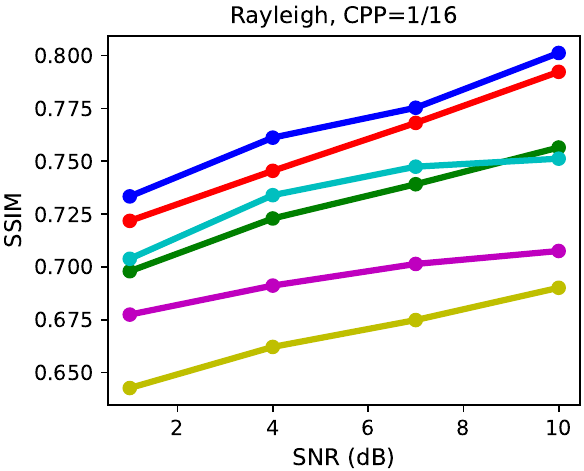}\\

\includegraphics[width=1.0\linewidth]{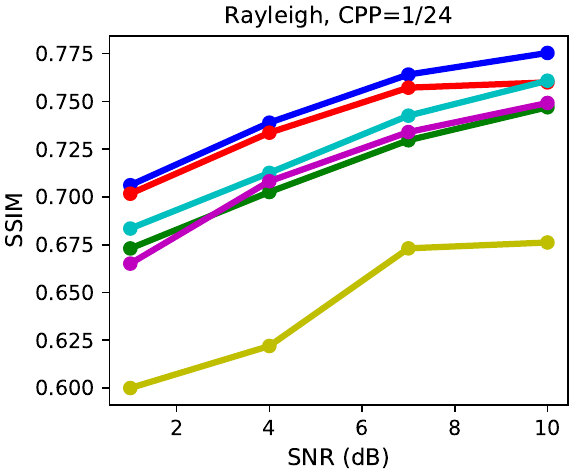}\\

\includegraphics[width=1.0\linewidth]{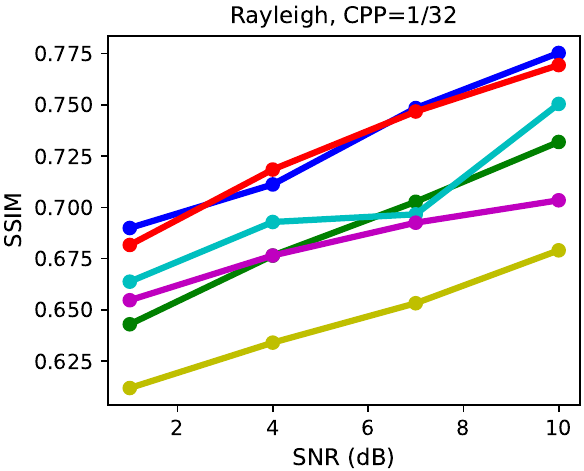}\\
\end{multicols}

\vspace{-0.15in}
\caption{SSIM results under different channel and CPP environments for DIV$2$K dataset.}\label{fig:ssim_result}
\vspace{-0.1in}
\end{figure*}

\begin{figure*}
\centering
\includegraphics[width=0.8\linewidth]{Figure3/plot_legend_main.pdf}\\
\begin{multicols}{4}
\centering
\includegraphics[width=1.0\linewidth]{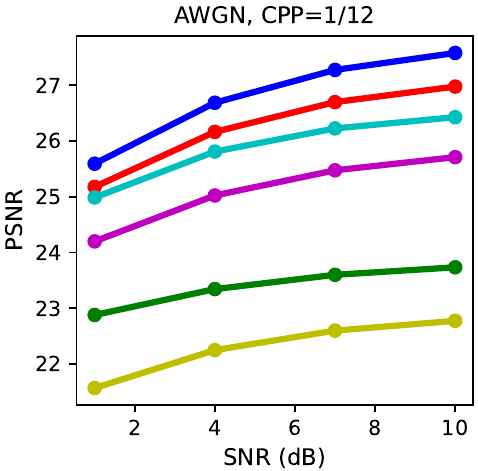}\\

\includegraphics[width=1.0\linewidth]{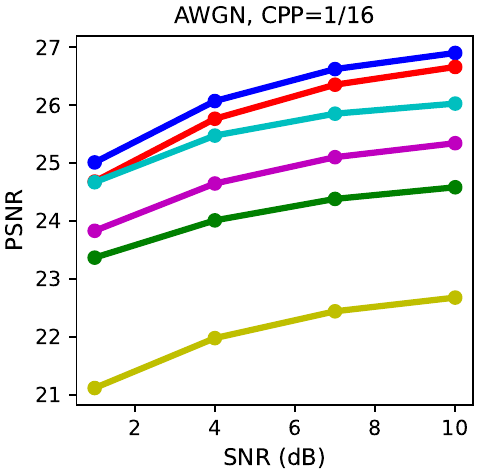}\\

\includegraphics[width=1.0\linewidth]{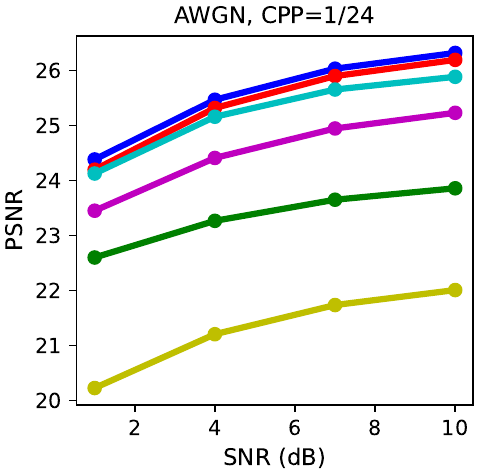}\\

\includegraphics[width=1.0\linewidth]{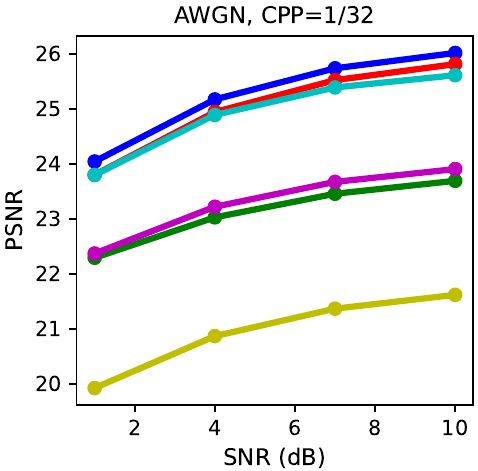}\\
\end{multicols}
\begin{multicols}{4}
\centering
\includegraphics[width=1.0\linewidth]{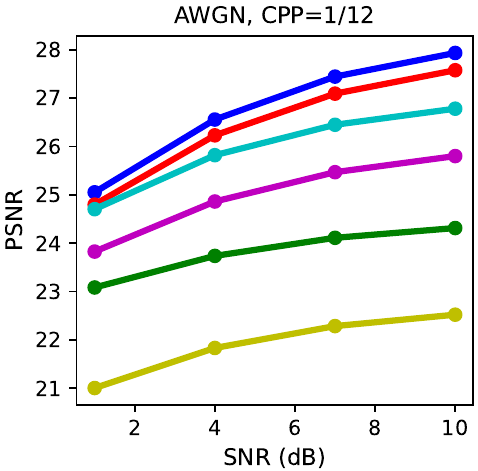}\\

\includegraphics[width=1.0\linewidth]{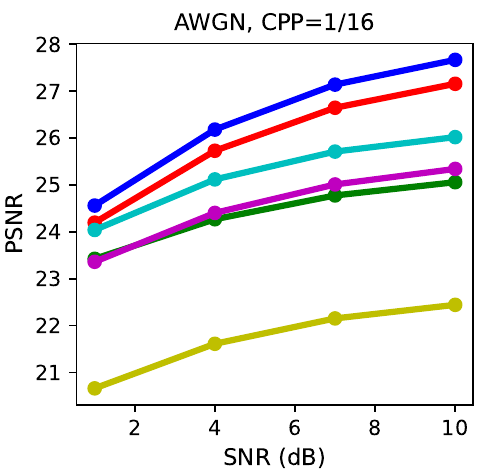}\\

\includegraphics[width=1.0\linewidth]{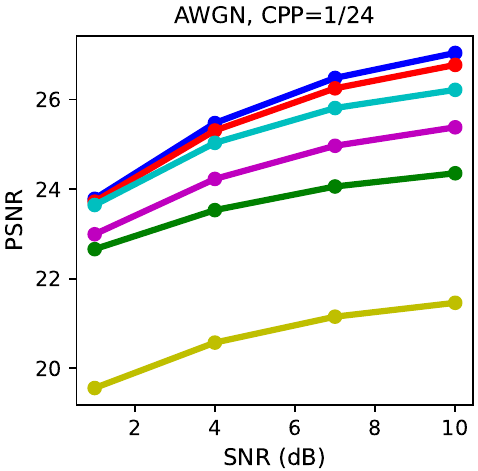}\\

\includegraphics[width=1.0\linewidth]{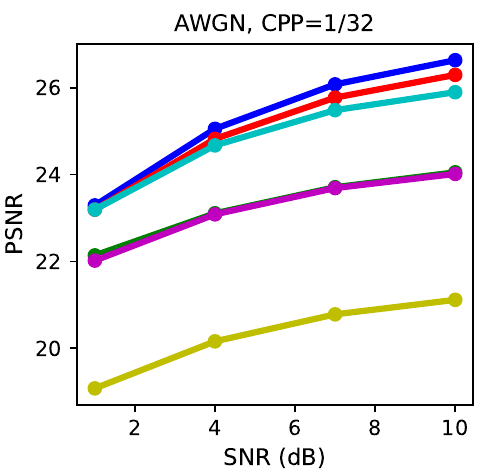}\\
\end{multicols}

\vspace{-0.15in}
\caption{PSNR results of models trained at fixed SNR for DIV$2$K dataset. The first row presents the results for models trained at an SNR of $4$ dB, while the second row corresponds to models trained at an SNR of $7$ dB.}\label{fig:fixSNR_result}
\vspace{-0.1in}
\end{figure*}

\begin{figure*}
\centering
\includegraphics[width=0.8\linewidth]{Figure3/plot_legend_main.pdf}\\
\begin{multicols}{4}
\centering
\includegraphics[width=1.0\linewidth]{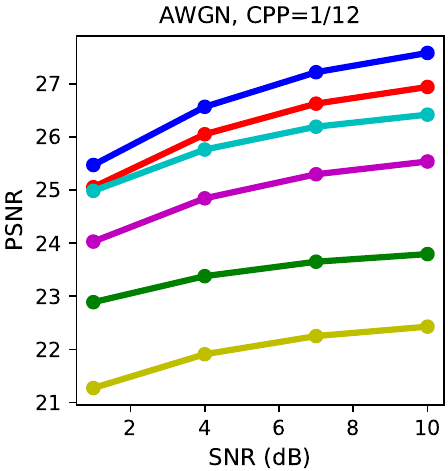}\\

\includegraphics[width=1.0\linewidth]{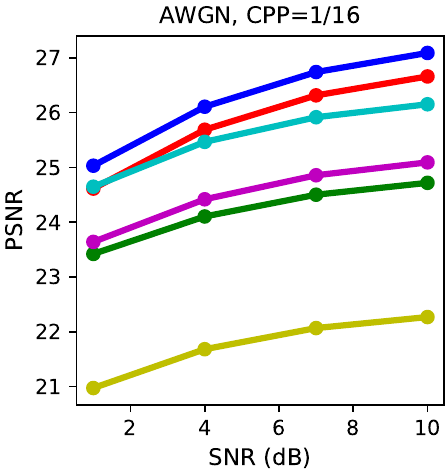}\\

\includegraphics[width=1.0\linewidth]{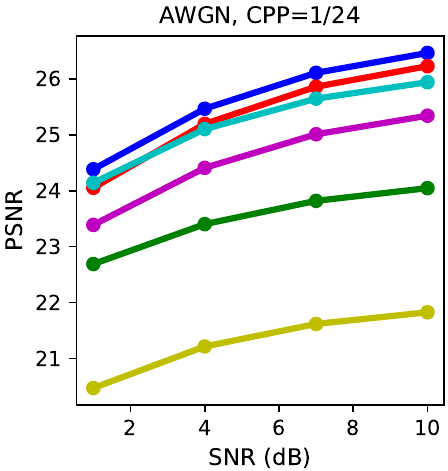}\\

\includegraphics[width=1.0\linewidth]{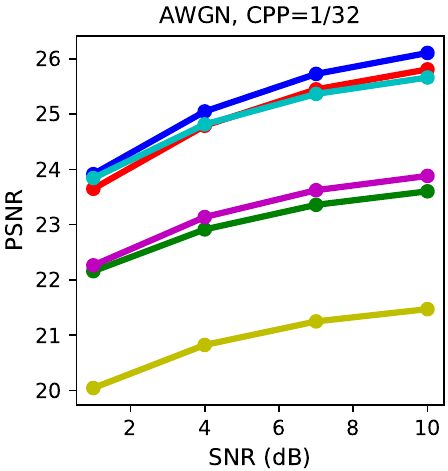}\\
\end{multicols}

\vspace{-0.15in}
\caption{PSNR results of models trained at randomly sampled SNRs from [$1,4,7,10$] for DIV$2$K dataset.}\label{fig:randomSNR_result}
\vspace{-0.1in}
\end{figure*}

\begin{figure*}
\centering
\includegraphics[width=0.8\linewidth]{Figure3/plot_legend_main.pdf}\\
\begin{multicols}{4}
\centering
\includegraphics[width=1.0\linewidth]{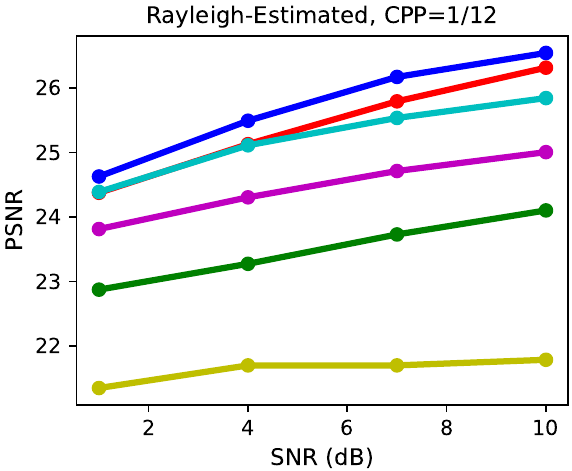}\\

\includegraphics[width=1.0\linewidth]{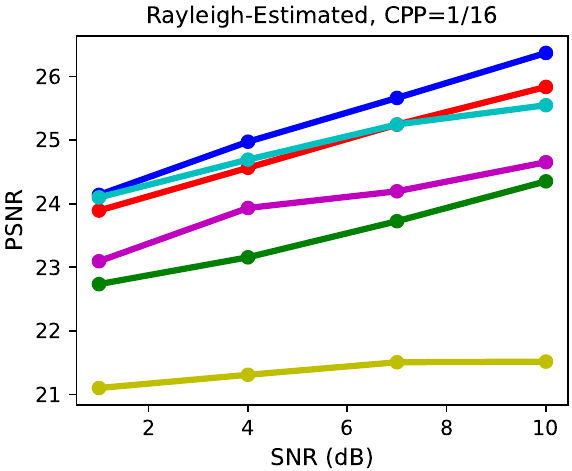}\\

\includegraphics[width=1.0\linewidth]{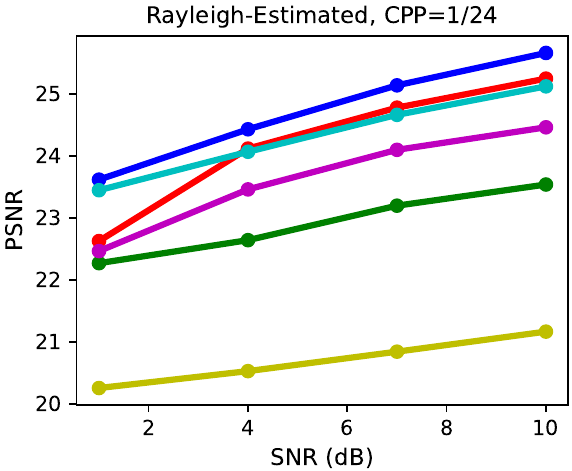}\\

\includegraphics[width=1.0\linewidth]{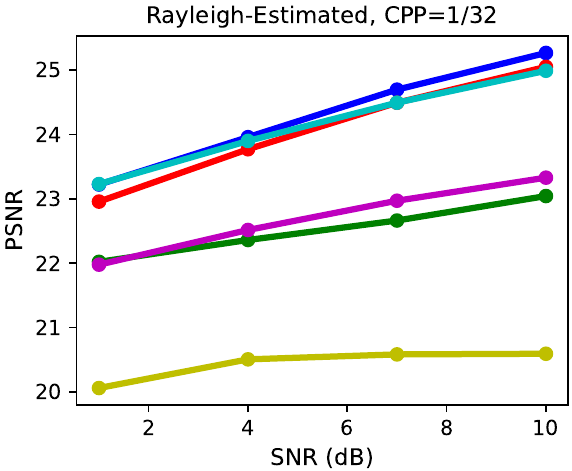}\\
\end{multicols}

\vspace{-0.15in}
\caption{PSNR results of models under the fast Rayleigh fading channel with estimated fading coefficients for DIV$2$K dataset.}\label{fig:Rayleigh_Estimated_result}
\vspace{-0.1in}
\end{figure*}

\begin{figure*}
\centering
\includegraphics[width=0.55\linewidth]{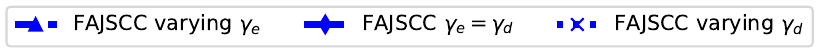}\\
\begin{multicols}{2}
\centering
\includegraphics[width=0.8\linewidth]{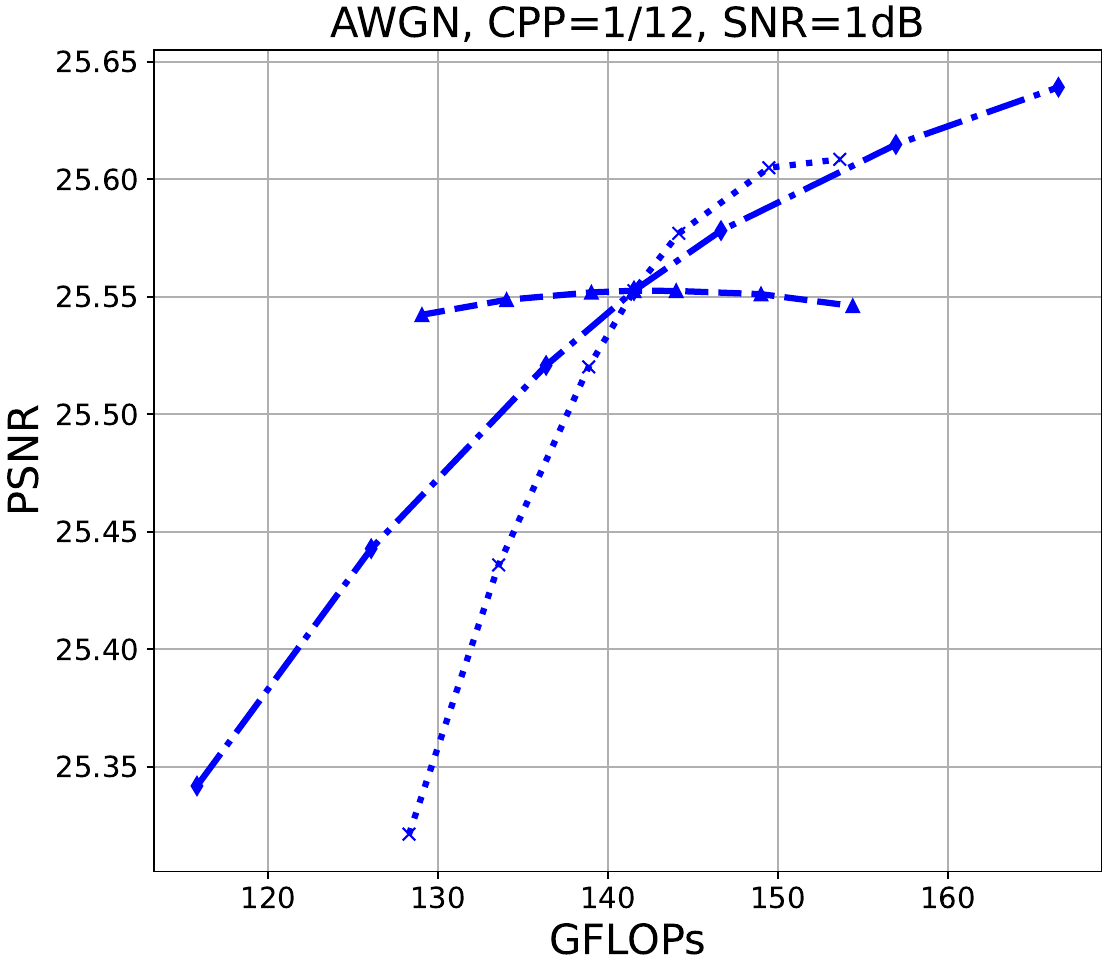}\\

\includegraphics[width=0.8\linewidth]{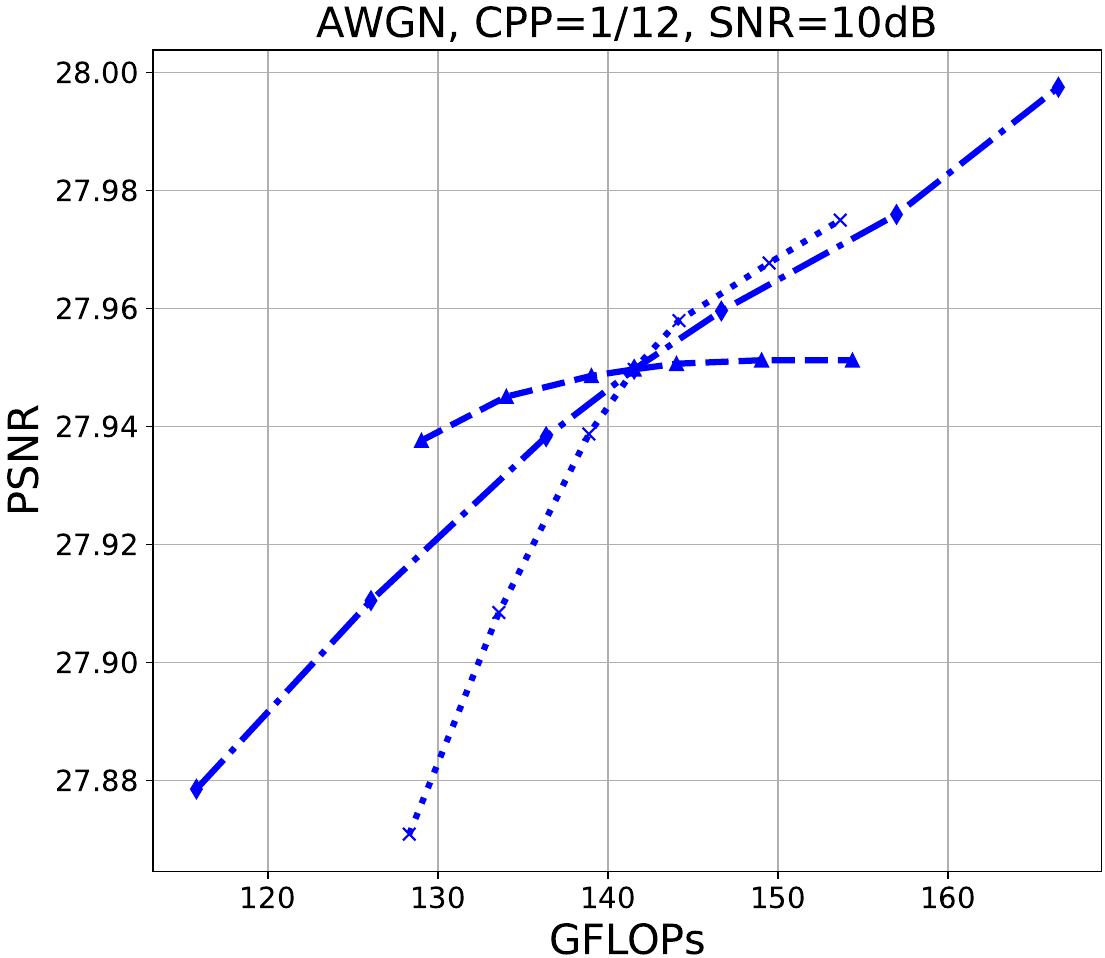}\\

\end{multicols}

\vspace{-0.15in}
\caption{Performance comparison for various importance ratios under different channel noises for DIV$2$K dataset. Varying $\gamma_{e}$: Only the importance ratio for the encoder’s FA blocks is changed. $\gamma_{e}= \gamma_{d}$: The importance ratios for the encoder's and decoder's FA blocks are changed together with the same value. Varying $\gamma_{d}$: Only the importance ratio for the decoder’s FA blocks is changed.}\label{fig:resource control}
\vspace{-0.1in}
\end{figure*}

\medskip
\noindent \textbf{PSNR and SSIM Results:} 
Figures~\ref{fig:psnr_result} and \ref{fig:ssim_result} show the PSNR and SSIM performance results across various communication environments. In these results, the models are trained at a fixed SNR and tested at the same SNR. Our LAJSCC and FAJSCC outperform the previous models for different performance metrics under various communication environments. This performance gain primarily comes from our dimension-specialized computation approaches: (1) Using only axis-dimension specialized operations (LAJSCC) shows higher performance than the recent SOTA SwinJSCC in most communication environments, while requiring computational resources about half and $14$ times smaller model storage size. This is a notable achievement, as even the recent SOTA lightweight model LICRFJSCC~\cite{yu2025novel} reduces computational resources by shrinking feature sizes when applying self-attention, which inherently limits its performance, making it impossible to reach that of SwinJSCC. (2) Upon the axis-dimension specialized operations, our proposed selective deformable self-attention (FAJSCC) even more enhances the image transmission performance by utilizing deformation and feature importance. These advantages lead FAJSCC to outperform SwinJSCC while using smaller computational and memory resources.

\medskip
\noindent \textbf{SNR Robustness:} Figure~\ref{fig:fixSNR_result} shows the results under a single-SNR training and multi-SNR testing scenario. The first row presents the results for models trained at an SNR of $4$ dB, while the second row corresponds to models trained at an SNR of $7$ dB. On the other hand, Figure~\ref{fig:randomSNR_result} shows the results of models trained at randomly sampled SNRs from [$1,4,7,10$]. In these results, our FAJSCC shows the best performance under various communication environments, and our LAJSCC shows better or comparable performance to previous deepJSCC models. Thus, our models are more robust than others under varying SNR environments.

\medskip
\noindent \textbf{Rayleigh Coefficient Robustness:}
In the above experiments on fast Rayleigh fading channels, we assumed perfect CSI only at the receiver side, i.e., the fast Rayleigh fading coefficient $\mathbf{h}$ is known. However, in real communication, the Rayleigh fading coefficient is estimated via a pilot signal. The relation between the $i$-th true Rayleigh fading coefficient $h_i$ and the $i$-th estimated Rayleigh fading coefficient $\hat{h}_i$ is as follows~\cite{capacity2006}.
\begin{align}
    \hat{h}_i = h_i + \psi_i, ~ \psi_i \sim \mathcal{CN}(0,\sigma_{\psi}^2),~ i \in [1:k]
\end{align}
where $\psi_i$ is a random estimation error. When $h_i \sim \mathcal{CN}(0,\sigma_{\psi}^2)$, current estimation technologies can achieve $\sigma_{\psi}^2=0.03$ to $\sigma_{\psi}^2=0.01$~\cite{estimator20022}. To evaluate the robustness of deepJSCCs for challenging constraints, we set $\sigma_{\psi}^2=0.03$, which can be achieved when the SNR of the pilot signal is $-5$ dB.

Figure~\ref{fig:Rayleigh_Estimated_result} presents the PSNR performance under the fast Rayleigh fading channel with MMSE equalization based on the estimated Rayleigh fading coefficient $\hat{\mathbf{h}}$. As observed in the figure, FAJSCC consistently achieves the best performance across varying SNRs and CPPs, showing robust superiority over the baseline models. LAJSCC also demonstrates strong performance and outperforms previous models under most SNR and CPP configurations.

\medskip
\noindent \textbf{Computation Resource Adjustment:}
Our FAJSCC is the first deepJSCC model capable of independently adjusting computational resources of the transmitter and receiver. In previous computational resource-adjustable deepJSCC models, the transceiver's computational complexity adjustments are constrained by the counterpart communication side~\cite{niu2025multimodal,zhang2024unified,raha2025dd}. As a result, such independent computational resource adjustment has not been possible in previous models. On the contrary, FAJSCC achieves this adjustment by independently varying the importance ratio $\gamma_{e}$ for the encoder’s FA blocks and $\gamma_{d}$ for the decoder’s FA blocks, without needing to agree with each other’s value. Since available computational resources for the encoder and decoder may vary over time, this flexible adjustment mechanism enhances the practicality of FAJSCC for real-world communication applications.

Figure~\ref{fig:resource control} illustrates PSNR performance versus GFLOPs as the importance ratios vary under different communication environments. The values of $\gamma_e$ and $\gamma_d$ range from $[0.0,0.2,0.4,0.5,0.6,0.8,1.0]$, and is fixed at $0.5$ when unspecified. Except for the case of varying $\gamma_{e}$ where curves are nearly flat, PSNR improves significantly as the importance ratio increases from $0$ to $0.5$. However, beyond $0.5$, the performance gain diminishes or even slightly decreases as the importance ratio approaches $1.0$. This indicates that less important features do not require extensive computational resources, validating our approach of selectively enhancing important features to increase computational efficiency. Moreover, this selective feature enhancement strategy does not have any information loss for adjusting computational complexity compared to the previous resource-adjustable deepJSCC methods that delete features~\cite{niu2025multimodal,zhang2024unified}. As a result, our FAJSCC maintains higher performance for various $\gamma$ than the current SOTA SwinJSCC, which shows $25.05$dB PSNR at SNR=$1$dB and $26.81$dB PSNR at SNR=$10$dB.

\medskip
\noindent \textbf{Complexity Demands at Encoder and Decoder:}
Recalling that deepJSCC involves the joint optimization of source coding and channel coding, we can further investigate the image reconstruction performance from three different perspectives. The first perspective is the encoder for source and channel coding where the extracted features should be both concise and robust to channel noise. The second perspective is the decoder for source coding where images are regenerated. The final perspective is the decoder for channel coding, which involves perceiving the meaning of the noisy received signal. By considering these components, our experiments with varying values of $\gamma_e$ and $\gamma_d$ reveal the relationship between computational complexity and the contribution of each component to reconstruction performance. In previous computational resource-adjustable deepJSCC models, the transceiver's computational complexity adjustments are constrained by the counterpart communication side~\cite{niu2025multimodal,zhang2024unified,raha2025dd}. As a result, such a discussion has not been possible in previous models.

First, when varying $\gamma_e$ (triangles in Figure~\ref{fig:resource control}), no significant change in PSNR is observed at both $\SNR=1$dB (low) and $\SNR=10$dB (high). This indicates that robust and concise feature extraction in the encoder does not require substantial computational resources, and adjusting the encoder’s computational budget has minimal impact on overall communication performance. On the other hand, when varying $\gamma_d$ (crosses in Figure~\ref{fig:resource control}), PSNR exhibits different behaviors depending on the SNR. The PSNR variation at $\SNR=1$dB is approximately $0.28$dB, which is about three times of the $0.09$dB variation at $\SNR=10$dB. Since the image regeneration tasks are the same at both SNRs, and thus the required computations are also the same, we can conclude that perceiving the meaning of the noisy received signal at the decoder demands significant computational complexity.

Our findings suggest that the decoder plays a more critical role than the encoder in deepJSCC performance, especially when SNR is low. This highlights the need to reconsider the conventional symmetric deepJSCC design, where the encoder and decoder typically have symmetric structures and computational resources~\cite{bourtsoulatze2019deep,yang2024swinjscc,wu2024mambajscc1}. We show that designing a larger decoder than the encoder is effective in the low SNR regime, even with a simple vanilla implementation, in the ablation studies at Section~\ref{sec:ablation}.

\begin{figure*}
    \centering
    \includegraphics[width=0.9\linewidth]{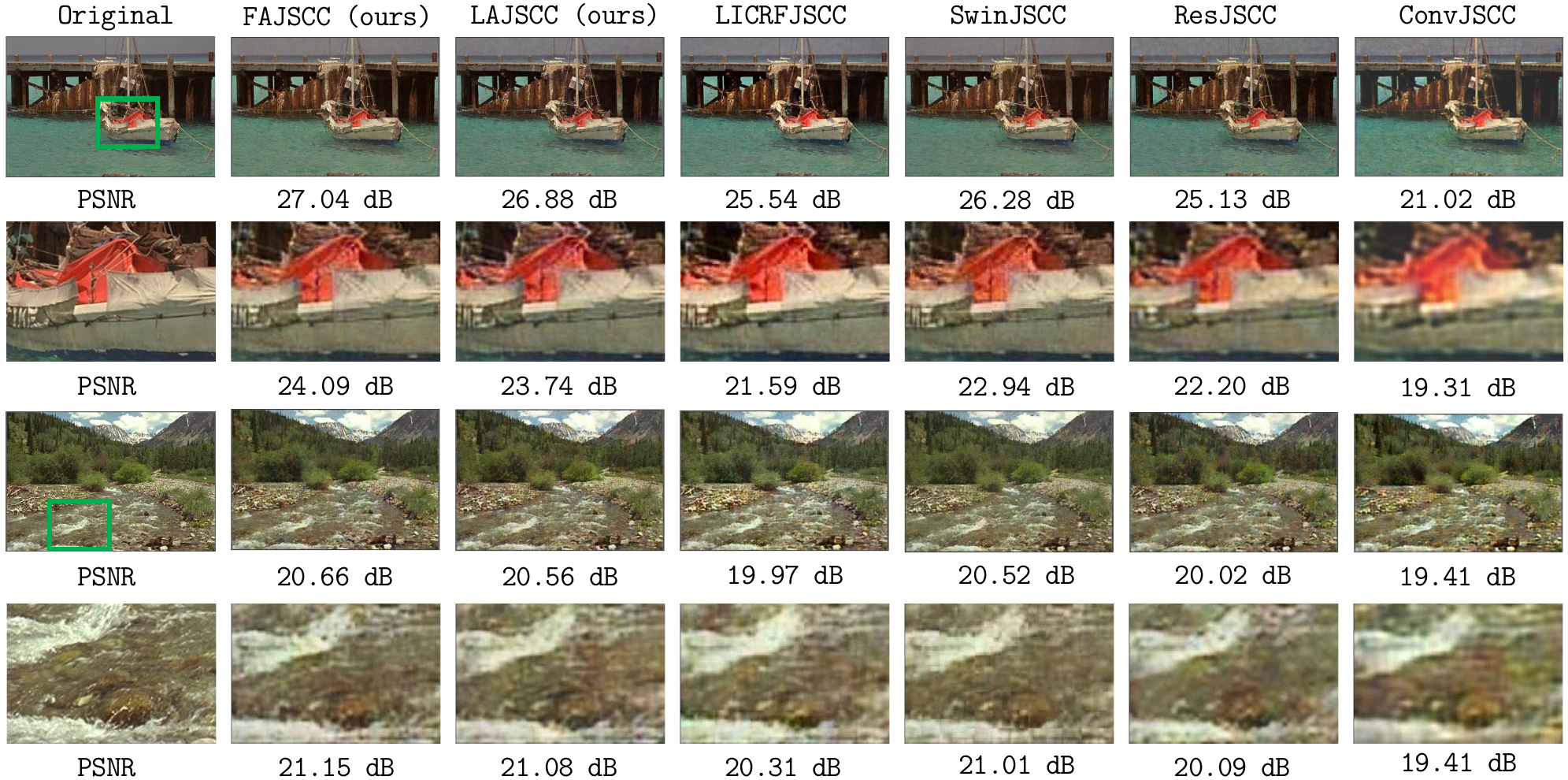}
    \caption{The first row shows the transmitted image at  $\CPP=\frac{1}{12}$, $\SNR=10$dB under the AWGN channel, and the third row shows the transmitted image at $\CPP=\frac{1}{12}$, $\SNR=1$dB under the AWGN channel for Kodak dataset. The second and fourth rows show the detailed results of the green box in the original images of the first and third rows.}\label{fig:visual result}
    \vspace{-0.15in}
\end{figure*}

\medskip
\noindent \textbf{Visual Inspection:} Figure~\ref{fig:visual result} visualizes two selected images from Kodak dataset. The first row shows a transmitted image at $\CPP=\frac{1}{12}$, $\SNR=10$dB under the AWGN channel, and the third row shows a transmitted image at $\CPP=\frac{1}{12}$, $\SNR=1$dB under the AWGN channel. The second and fourth rows provide zoomed-in views of the green box in the original images of the first and third rows. 

As illustrated in the figure, FAJSCC and LAJSCC reconstruct image details with fewer artifacts and greater clarity compared to previous deepJSCC methods. For example, in the second row, ours can reconstruct the ship's sail more clearly than others. In the fourth row, ours can reconstruct a more detailed water hole than others. These qualitative results further validate the superior image transmission capabilities of our proposed models, aligning with the quantitative performance results.

\begin{table*}[t]
\centering
\caption{The ablation study for verifying the effectiveness of our proposed methods and deepJSCC complexity analysis under AWGN channel and CPP=$1/32$. PSNR ($\mathrm{dB}$) values of each SNR, and GFLOPs and Memory (MB) usages are evaluated for DIV$2$K dataset.}
\label{tab:ablation_study}
\begin{tabular}{|l||c|c|c|c|c|c|c|c|}
\hline
& \makecell{PSNR at \\ SNR=$1$dB} & \makecell{PSNR at \\ SNR=$4$dB} & \makecell{PSNR at \\ SNR=$7$dB} & \makecell{PSNR at \\ SNR=$10$dB} & \makecell{Latency \\ (ms/image)} & GFLOPs & \makecell{Peak Memory \\ (MB)} & \makecell{Model Storage Size \\ (MB)}\\ \hline\hline
SwinJSCC~\cite{yang2024swinjscc} & $23.86${\tiny $\pm 0.12$} & $24.80${\tiny $\pm 0.06$} & $25.45${\tiny $\pm 0.08$} & $25.73${\tiny $\pm 0.20$} & $704.51${\tiny $\pm 2.57$}      & $160.04$  & $3347.44$       & $72.24$  \\ \hline
LAJSCC                           & $23.86${\tiny $\pm 0.05$} & $24.86${\tiny $\pm 0.11$} & $25.70${\tiny $\pm 0.08$} & $26.55${\tiny $\pm 0.13$} & $228.25${\tiny $\pm 1.02$}      & $90.66$   & $1421.05$       & $5.31$   \\ \hline
FAJSCC                           & $24.01${\tiny $\pm 0.05$} & $25.09${\tiny $\pm 0.07$} & $26.08${\tiny $\pm 0.03$} & $26.82${\tiny $\pm 0.15$} & $355.74${\tiny $\pm 0.63$}      & $140.16$  & $2122.69$       & $13.62$  \\ \hline\hline
FAJSCC w/ PGB                    & $24.00${\tiny $\pm 0.03$} & $25.11${\tiny $\pm 0.05$} & $26.08${\tiny $\pm 0.03$} & $26.85${\tiny $\pm 0.11$} & $355.68${\tiny $\pm 1.07$}      & $140.17$  & $2122.67$       & $13.62$  \\ \hline
FAJSCC w/o AT                    & $23.87${\tiny $\pm 0.08$} & $25.04${\tiny $\pm 0.11$} & $26.11${\tiny $\pm 0.11$} & $26.96${\tiny $\pm 0.10$} & $392.89${\tiny $\pm 1.71$}      & $151.08$  & $2102.71$       & $14.49$  \\ \hline\hline
FAJSCC ($s=0.3$)& $24.00${\tiny $\pm 0.05$} & $25.07${\tiny $\pm 0.07$} & $26.06${\tiny $\pm 0.03$} & $26.80${\tiny $\pm 0.15$} & $355.72${\tiny $\pm 0.61$}      & $140.16$  & $2122.69$       & $13.62$  \\ \hline
FAJSCC ($s=0.6$)& $23.97${\tiny $\pm 0.05$} & $25.04${\tiny $\pm 0.06$} & $26.01${\tiny $\pm 0.03$} & $26.76${\tiny $\pm 0.15$} & $355.72${\tiny $\pm 0.61$}      & $140.16$  & $2122.69$       & $13.62$  \\ \hline
FAJSCC ($s=1.0$)& $23.93${\tiny $\pm 0.05$} & $24.92${\tiny $\pm 0.06$} & $25.94${\tiny $\pm 0.03$} & $26.68${\tiny $\pm 0.15$} & $355.72${\tiny $\pm 0.61$}      & $140.16$  & $2122.69$       & $13.62$  \\ \hline\hline
FAJSCC w/o FD                    & $24.09${\tiny $\pm 0.02$} & $25.10${\tiny $\pm 0.07$} & $25.88${\tiny $\pm 0.12$} & $26.71${\tiny $\pm 0.10$} & $319.28${\tiny $\pm 0.74$}      & $134.92$  & $1878.25$       & $13.32$  \\ \hline
FAJSCC w/o LA                    & $23.98${\tiny $\pm 0.04$} & $24.97${\tiny $\pm 0.06$} & $25.90${\tiny $\pm 0.09$} & $26.72${\tiny $\pm 0.14$} & $341.28${\tiny $\pm 1.16$}      & $140.02$  & $2142.38$       & $13.11$  \\ \hline\hline
LAFAJSCC                         & $24.06${\tiny $\pm 0.02$} & $25.09${\tiny $\pm 0.07$} & $25.98${\tiny $\pm 0.08$} & $26.70${\tiny $\pm 0.14$} & $310.48${\tiny $\pm 1.28$}      & $120.61$  & $2232.87$       & $9.34$   \\ \hline
FALAJSCC                         & $23.88${\tiny $\pm 0.07$} & $24.94${\tiny $\pm 0.08$} & $25.82${\tiny $\pm 0.03$} & $26.65${\tiny $\pm 0.06$} & $306.34${\tiny $\pm 1.34$}      & $120.57$  & $2227.93$       & $9.34$   \\ \hline
\end{tabular}
\end{table*}

\subsection{Ablation Studies}
\label{sec:ablation}
The result of the top three models in Table~\ref{tab:ablation_study} shows that our proposed axis dimension specialized processing (LAJSCC) is sufficient to surpass the current SOTA SwinJSCC~\cite{yang2024swinjscc} with much smaller deployment costs. Moreover, using a selective deformable self-attention (FAJSCC) further improves image transmission performance for various SNR settings. In addition, we conduct extensive ablation studies, examining each component point by point to verify the effectiveness of our proposed methods.

\medskip
\noindent \textbf{Attention Feature Extraction:} 
To verify the effectiveness of our proposed attention family tree structure, we compare two models, which are variants of our FAJSCC: FAJSCC with partial gradient block (FAJSCC w/ PGB) and FAJSCC without attention family tree (FAJSCC w/o AT). The architecture of FAJSCC w/ PGB is the same as the FAJSCC, but we prevent backward-gradient flow from the window importance head to the spatial importance in the attention family tree of Figure~\ref{fig:dimension_specialized} during training. Since the spatial importance branch only learns to extract information for spatial attention in this training setting, the window importance head should predict window importance based only on the spatial attention information. The results show that FAJSCC w/ PGB performance is very similar to FAJSCC. This means that spatial attention and window importance are highly related, so that information for extracting spatial attention is sufficient to extract window importance. Thus, our attention family tree structure extracts each attention feature efficiently by removing redundant overlapping processing steps based on the relations of offset, attention feature, and window importance.

Another direction of the ablation study also proves the efficiency of our attention family tree. FAJSCC w/o AT is constructed by replacing the attention family tree with the independent parallel predictors for each attention feature, offset, and window importance. The results show that FAJSCC shows better or similar performance compared to FAJSCC w/o AT with smaller computational burden and latency. More precisely, FAJSCC ($0.181$ dB/GFLOPs) has $7.73 \%$ improved PSNR/GFLOPs than FAJSCC w/o AT ($0.168$ dB/GFLOPs) on average. In the case of latency, FAJSCC ($0.071$ dB/ms) has $10.93 \%$ improved PSNR/latency than FAJSCC w/o AT ($0.064$ dB/ms) on average. These efficiency gains verify the design of our attention family tree.

Compared to FAJSCC w/o AT, FAJSCC exhibits both disadvantages and advantages. (1) Disadvantage: The branches in the proposed attention family tree are required to extract shared features that support multiple attention variants. This structural constraint may limit the specialization capability of each branch compared to the independent parallel predictors in FAJSCC w/o AT. (2) Advantage: Shared representations across multiple branches can be interpreted as a form of parameter sharing, analogous to multi-task learning. It has been shown that shared features reduce overfitting and improve generalization by limiting model variance~\cite{ruder2017overview}. In noisy channel conditions, especially at low SNR, where input perturbations are dominant, such variance reduction may alleviate sensitivity to channel noise. Consequently, FAJSCC achieves superior performance in low-SNR regimes. However, when the SNR is high and channel noise becomes negligible, the regularization effect of shared representations may introduce slight bias, leading to marginally inferior performance compared to the fully independent predictors.

\medskip
\noindent \textbf{Importance Robustness:} 
Our FAJSCC relies on importance estimation for selective feature enhancement. To evaluate its robustness, we randomly swap a portion of important and less important features during inference. In Table~\ref{tab:ablation_study}, FAJSCC ($s$) denotes the model where a fraction $s$ of important features are replaced with less important ones. When $s$ is moderate, the performance degradation is minimal, demonstrating the robustness of FAJSCC. This robustness arises from the independent importance estimation in each FA block, which allows important features to be selectively enhanced in other blocks even if they are misidentified in a particular block. In contrast, when all important features are replaced ($s=1.0$), the performance drops significantly, confirming the necessity of selective feature enhancement for important features.

\medskip
\noindent \textbf{Attention Mechanism:} 
To verify the effectiveness of our feature deformation for selective self-attention, we make a new variant of FAJSCC that does not use feature deformation (FAJSCC w/o FD). The results show that FAJSCC is better than FAJSCC w/o FD in most SNR settings, which validates the need to use feature deformation before self-attention as we proposed. We also checked whether using spatial attention and channel attention mechanisms enhances feature processing ability by making a new variant of FAJSCC that does not use these lightweight attentions (FAJSCC w/o LA). The results show that FAJSCC is better than FAJSCC w/o LA in most SNR settings, which validates the need to use each lightweight attention before depthwise convolution and pointwise convolution as we proposed.

Our proposed feature enhancement methods improve performance by mitigating degradations caused by the limited power and bandwidth of the communication channel. Such degradations could be further alleviated by developing restoration methods for transmitted images, such as modern deblurring techniques~\cite{zhang2018adversarial,zhang2023mc,zhang2022deep}.

\medskip
\noindent \textbf{Complexity Demands of DeepJSCC:} In Section~\ref{sec:main}, we showed that perceiving the meaning of the noisy received signal requires the highest computational resources; thus, an asymmetric structure that takes a larger decoder than the encoder will have high efficiency when SNR is low. We show this by comparing three models: 1) FAJSCC, 2) LAFAJSCC where the encoder is from LAJSCC and the decoder is from FAJSCC, and 3) FALAJSCC where the encoder is from FAJSCC and the decoder is from LAJSCC. The results show that LAFAJSCC is better or comparable to the FAJSCC in low SNR regimes while using lower computational resources. Moreover, FALAJSCC is worse than LAFAJSCC in low SNR regimes while using similar computational resources. This highlights the importance of adapting the encoder–decoder asymmetry to the channel condition. Therefore, utilizing computational complexity demands under the given SNR will lead to highly efficient communication systems.

Although it is well known that the decoding complexity of traditional channel coding schemes (e.g., LDPC, turbo codes) increases significantly with higher error probabilities, in contrast to their near-constant encoding time~\cite{shao2019survey}, the computational complexity requirements of deepJSCC have not been thoroughly investigated. In this work, we demonstrate how performance varies according to encoder and decoder computational complexity, respectively, in Figure~\ref{fig:resource control}. Moreover, the results in Table~\ref{tab:ablation_study} indicate that allocating greater computational resources to the decoder can improve performance than to the encoder under low-SNR conditions. A more formal analysis of the computational resource requirements of encoders and decoders across different channel environments is left for future work.

\subsection{Comparison with separation-based methods}
\label{sec:separation}

\begin{table*}
\centering
\caption{Bjøntegaard delta (BD) metric comparison (PSNR) between modern communication systems and FAJSCC over an AWGN channel at SNR = 10 dB on the Kodak dataset.}
\label{tab:separation_system}
\begin{tabular}{|l||l|l|l|l|}
\hline
             & BD-CPP ($\%$)                & BD-PSNR (dB) & Latency (sec/image) & BD-PSNR/Latency (dB/sec) \\ \hline
JPEG         & 63.05                        & -2.28        & 0.01              & N/A                      \\ \hline
JPEG2000 (base)     & 0.00                         & 0.00         & 1.06              & N/A                      \\ \hline
BPG          & -34.63                       & 1.97         & 0.63              & 3.12                     \\ \hline
VTM          & -49.24                       & 3.02         & 58.20             & 0.05                     \\ \hline
huge FAJSCC  & -32.48{\tiny $\pm 0.54$}     & 1.67{\tiny $\pm 0.08$}           & 0.24              & 6.95                     \\ \hline
\end{tabular}
\end{table*}

\begin{table}
\centering
\caption{Bjøntegaard delta (BD) metric comparison (MS-SSIM) between modern communication systems and FAJSCC over an AWGN channel at SNR = 10 dB on the Kodak dataset.}
\label{tab:separation_system_msssim}
\begin{tabular}{|l||l|l|l|l|}
\hline
             & BD-CPP ($\%$) & BD-MS-SSIM (dB)\\ \hline
JPEG         & 22.09         & -0.98\\ \hline
JPEG2000 (base)     & 0.00          & 0.00 \\ \hline
BPG          & -42.11        & 2.24 \\ \hline
VTM          & -54.03        & 3.12 \\ \hline
huge FAJSCC  & -63.94{\tiny $\pm 0.68$}        & 3.95{\tiny $\pm 0.12$} \\ \hline
\end{tabular}
\end{table}

\begin{table}
\centering
\caption{Architecture Summary Table for huge FAJSCC.}
\label{tab:hugearc_summary}
\begin{tabular}{|l||l|}
\hline
The number of stages ($L$)                & $4$ \\ \hline
The number of FA blocks per stage ($n_i$) & $[2,2,6,2]$ \\ \hline
Feature channel sizes ($C_i$)     & $[128, 192, 256, 320]$ \\ \hline
Window sizes                      & $[8,8,8,8]$ \\ \hline
The number of attention heads             & $[1,1,1,1]$ \\ \hline
\end{tabular}
\end{table}

To contextualize our contributions, we compare our FAJSCC with modern separation-based methods under an AWGN channel at SNR=$10$. Modern separation-based methods consist of source coding, channel coding, and modulation schemes. Following previous research~\cite{bourtsoulatze2019deep}, we use $2/3$-rate LDPC codes for channel coding and $16$-QAM for modulation, i.e., the best setting under the AWGN channel at SNR=$10$. By varying source coding methods from old JPEG to recent VTM, we compare our FAJSCC with various separation-based methods. We use the standard source coding results from the image compression research framework~\cite{begaint2020compressai}.\footnote{https://github.com/InterDigitalInc/CompressAI/tree/master/results/image/kodak.} These source coding results were obtained by using $80$ Intel Xeon Gold $5218$R CPU with $2.10$GHz. Since the source coding results do not report exact values for the four CPP settings $\left[\frac{1}{12}, \frac{1}{16}, \frac{1}{24}, \frac{1}{32}\right]$, we use the closest available points corresponding to these CPP values.

\medskip
\noindent \textbf{Experimental Setup:}
We employed the Bjøntegaard delta (BD) metrics to quantify the performance gains~\cite{barman2024bj}. In Table~\ref{tab:separation_system}, BD-CPP ($\%$) denotes the percentage reduction in CPP required to achieve the same PSNR as the anchor model, whereas BD-PSNR (dB) represents the average PSNR improvement over the anchor model at an equivalent CPP. In Table~\ref{tab:separation_system_msssim}, BD-CPP ($\%$) denotes the percentage reduction in CPP required to achieve the same Multiscale structural similarity (MS-SSIM) index as the anchor model, whereas BD-MS-SSIM (dB) represents the average MS-SSIM improvement over the anchor model at an equivalent CPP. Here, MS-SSIM extends SSIM by computing the structural similarity at multiple image resolutions and combining them, so that both fine details and large-scale structural information are jointly evaluated instead of being measured at a single scale. We set JPEG2000 as the anchor model in Table~\ref{tab:separation_system}. 

To obtain BD-metrics, the CPP and performance metric regions should overlap across the baselines. Since our lightweight FAJSCC has only a small overlap with recent source coding methods, we train huge FAJSCCs to obtain valid BD-metrics. The architecture is summarized in Table~\ref{tab:hugearc_summary}. Our huge FAJSCC is trained on a Flickr$30$k dataset~\cite{plummer2015flickr30k} with a batch size of $8$ for $80$ epochs. During training, the images are randomly resized and cropped to $256 \times 256$. The training takes about $4$ days for each huge FAJSCC model. Owing to the considerable training time, the same huge FAJSCC model trained under $\mathcal{L}_{\distortion}=\text{MSE}$ setting was used when evaluating both PSNR and MS-SSIM. The other settings are the same as the FAJSCC.

\medskip
\noindent \textbf{Performance Comparison:}
As shown in Tables~\ref{tab:separation_system} and \ref{tab:separation_system_msssim}, the performance trends depend on the distortion metric considered. Under the PSNR criterion, huge FAJSCC achieves competitive BD-CPP and BD-PSNR performance compared with modern codecs. Although VTM provides the highest BD-PSNR gain, FAJSCC attains a comparable CPP–distortion efficiency to BPG while maintaining substantially lower latency than both BPG and VTM. Our huge FAJSCC ($6.95$ dB/sec) shows significantly higher BD-PSNR/Latency efficiency compared to modern BPG ($3.12$ dB/sec) and VTM ($0.05$ dB/sec).

When evaluated using MS-SSIM, FAJSCC demonstrates a clear performance advantage. As reported in Table~\ref{tab:separation_system_msssim}, FAJSCC achieves the largest BD-CPP reduction ($-63.94\%$) and the highest BD-MS-SSIM gain ($3.95$ dB), outperforming both BPG and VTM. These results indicate that FAJSCC is particularly effective under structural similarity accuracy.

It is also noteworthy that the results for BPG and VTM were obtained using a 2020 high-end CPU (Intel Xeon Gold 5218R), whereas huge FAJSCC was evaluated on a 2016 high-end GPU (NVIDIA Tesla P100). Therefore, the latency advantage of FAJSCC is expected to become even more significant in practical deployment scenarios. Considering both perceptual quality gains and latency efficiency, FAJSCC offers a compelling solution for IoT applications requiring fast and high-quality image transmission, such as real-time monitoring systems. These findings further support the practical necessity of FAJSCC over conventional digital communication frameworks.

\section{Conclusion}\label{sec:conclusion}
In this paper, we introduced FAJSCC, a novel framework for efficient image transmission with adjustable computational complexity based on the attention family tree. Leveraging axis-dimension specialized computation, FAJSCC reduces computational cost while preserving feature expressiveness. Building on this foundation, our selective deformable self-attention mechanism further enhances transmission performance by adaptively refining the most important features. In addition, our attention family tree efficiently provides attention features for axis-specialized computation and selective deformable self-attention by eliminating overlapping computations. The flexibility of selective deformable self-attention ensures strong performance under varying computational constraints, consistently outperforming SwinJSCC. Moreover, FAJSCC is the first deepJSCC model capable of independently adjusting the computational resources of the encoder and decoder. Our performance analysis reveals that the decoder's noisy received signal perception function consumes the largest fraction of computational complexity. A promising future direction is to develop an improved FAJSCC that dynamically allocates computational resources to the decoder's signal perception module based on channel conditions while minimizing unnecessary computation on the encoder side. This would maximize deepJSCC’s efficiency and adaptability in real-world communication systems.

\end{document}